\documentclass[11pt,a4paper]{article}

\usepackage{fullpage}
\usepackage{algorithm}
\usepackage{algpseudocode}
\usepackage{hyperref}
\usepackage{placeins}
\usepackage[T1]{fontenc}
\usepackage[utf8]{inputenc}
\usepackage{authblk}
\usepackage{natbib}
\usepackage{graphicx}
\usepackage{subcaption}
\usepackage{booktabs}
\usepackage{amssymb}

\title{Location- and time-dependent meeting point recommendations for shared interurban rides}

\author[1]{Paul Czioska\thanks{paul.czioska@ikg.uni-hannover.de}}
\author[2]{Aleksandar Trifunovi{\'c}\thanks{a.trifunovic@tu-braunschweig.de}}
\author[3]{Sophie Dennisen\thanks{sophie.dennisen@tu-clausthal.de}}
\author[1]{Monika Sester\thanks{monika.sester@uni-hannover.de}}

\affil[1]{Institute of Cartography and Geoinformatics, Leibniz Universit\"at Hannover, Germany}
\affil[2]{Institute of Transportation and Urban Engineering, Technische Universit\"at Braunschweig, Germany}
\affil[3]{Department of Informatics, Technische Universit\"at Clausthal, Germany}

\begin{document}

\maketitle

\begin{abstract}
Drivers offering spare seats in their vehicles on long-distance (interurban) trips often have to pick up or drop off passengers in cities en route. In that case it is necessary to agree on a meeting point. Often, this is done by proposing well-known locations like train stations, which frequently induces unnecessary detours through the inner-city districts. In contrast, meeting points in the vicinity of motorways and arterial roads with good public transport connection can reduce driving time and mileage. This work proposes a location-based approach to enable a fast and automatic recommendation of suitable pick-up (and drop-off) points for drivers and passengers using a GIS workflow and comprehensive precomputation of travel times.
\end{abstract}


\textit{Preprint submitted to Journal of Location Based Services}

% CONTENT
\section{Introduction}
\label{sec:Introduction}

% Background: What is the overall problem?
Nowadays the society is facing a large demand for individual mobility. Private vehicles still play a major role in satisfying this demand, resulting in congested streets and environmental pollution. Low occupancy of private cars is one of the reasons for the large number of vehicles on the streets. In Germany, for example, the average car occupancies range from 1.1 average people per vehicle for daily commuting trips to 1.9 for leisure trips \citep{infas2010mobilitat}. In the US, only 9.2 \% of the commuters carpool \citep{dot2015national}. Hence, it is reasonable to increase the occupancy of the vehicles by sharing rides in order to reduce the car traffic.

% Motivation: WHY is it relevant?
Long-distance (interurban) ride-sharing focuses on occasional trips, mostly with scheduling in advance and no strict requirements of meeting times \citep{furuhata2013ridesharing}. It is a frequently used way of traversing long distances and an alternative to trains and intercity bus services with the possibility to share the travel costs among all passengers. In recent years, many online ride-matching platforms have evolved with the goal of matching travelers with similar itineraries, like BlaBlaCar\footnote{\url{http://www.blablacar.com}} or Liftshare\footnote{\url{http://www.liftshare.com}}.

Drivers offering rides on ride-sharing platforms often find passengers who need to be picked up or dropped off in cities en route. Hence, it is necessary to negotiate a meeting (and/or drop-off) point. For this purpose, common locations that are well-known, easy to describe and well reachable by public transport are frequently chosen, e.g. the central train or bus station or salient landmarks. However, such locations are usually located in the inner-city districts, inducing unnecessary detours and time loss for the drivers. In contrast, the use of meeting points close to motorways or arterial roads which are further easily reachable by public transport could reduce driving time, driving distance and congestion in urban areas. A recommendation of such points is especially helpful when a driver or a passenger is not familiar with the environment. The problem is well known in the community, and there are already some solutions available: recently, for example, BlaBlaCar started a service allowing drivers to choose a meeting point from a set of predefined candidate locations for some cities when creating a ride \citep{BlaBlaCarMP}. However, the set of candidates is often very limited, and the passengers have no influence on the meeting point selection. Hence, other meeting points than the selected one could be more time efficient or beneficial from other points of view for the passengers.

% State-of-the-art: WHAT has been done so far?
A ride-sharing system is, in the broader sense, defined as a system to bring together travellers with similar itineraries and time schedules \citep{agatz2011dynamic}. The ride-sharing problem aims at coordinating a certain demand and a number of available vehicles to create feasible matches subject to various constraints, such as travel time or vehicle capacity limitations. It can be classified into various groups of different usage \citep{furuhata2013ridesharing}. Numerous optimization strategies have been developed to solve the matching of drivers to passengers in different constellations \citep{agatz2012optimization}. A basic differentiation can be done between the static and the dynamic approach. In the static case all requests are known in advance, but in the dynamic case rides can be announced on short notice (ad-hoc).

In contrast to ride-sharing systems with a pick-up (and drop-off) directly at the origin (and destination, respectively), the usage of meeting points has not gained much attention in the literature. If the driver/passenger pairs are fixed, intermediate meeting locations are necessary to indicate which part of the trip can be traveled together \citep{aissat2014dynamic, aissat2015meeting}. Relaxing the constraint of fixed pairs further increases the degree of complexity. \cite{balardino2016heuristic} propose a greedy and an iterated local search heuristic to assign passengers to drivers at meeting points (called \emph{Close Enough Points}), given that the possible driver detour and the vehicle capacity is limited. In the work of \cite{stiglic2015benefits}, potential benefits of meeting points in a ride-sharing system are investigated by a computational study using randomly distributed meeting points in the Euclidean plane. They show that the introduction of meeting points can improve several metrics like the percentage of matched participants or saved vehicle mileage. 

Further, there are some approaches to ease the meeting point determination for the passengers. \cite{Rigby:2013:OCU:2525314.2525334} developed an opportunistic user interface application using so called \emph{launch pads}, which help passengers to find a ride by visualizing the area in which they could potentially be picked up. The idea is based on the principles of time-geography and space-time-prisms \citep{hagerstraand1970people,miller1991modelling}, a commonly applied technique for dynamic ride-sharing models \citep{raubal2007time,winter2006ad}. Later, the launch pad idea was extended by an enhanced 3D visualization to improve the decision making process, demonstrating the importance of human-computer interactions in a ride-sharing scenario \citep{rigby2015enhancing}. In addition, the concept has the ability to preserve the user privacy, since a contract can be established without a need for revealing the actual origin. In fact, privacy can be seen as a further major advantage when using meeting points, since there are some techniques available to control the own data of the ride sharing users \citep{aivodji2016meeting,goel2016privacy}.

% Knowledge gaps: WHAT is currently missing?
However, most existing approaches focus on intra-urban rides covering shorter distances, where the reachable meeting points for passengers are limited by a walking threshold. In this paper we aim to extend the meeting point search by including public transportation, allowing the passengers to reach more remote meeting points, e.g. close to motorway exits, which is most relevant if the driver is intending to just pass the city. In a prospective recommender system application, the results should be available in real-time. Since every meeting point recommendation is based on an optimization procedure with increasing complexity for increasing number of participants, response time is a significant factor. Hence, we propose an extensive precomputation of shortest paths to substantially reduce the query time. This technique is often applied for route-planning algorithms, e.g. for distance tables in hierarchical routing networks \citep{sanders2005highway} or precomputed cluster distances \citep{maue2009goal}.

% Objectives: What is the paper doing?
In this paper we propose a location-based method to recommend real-world meeting points to long-distance ride-sharing customers. In our scenario we assume a \textit{driver} passing the city on major roads, having to pick-up one or multiple \textit{passengers} at exactly one point in the city. While the driver is supposed to drive on the streets, the passengers are supposed to walk and use the public transport to reach the meeting point. The goal is to determine the most beneficial location among a set of predefined candidate locations, based on the current spatio-temporal location of driver and passengers. The recommended meeting point should minimize the total or maximum time consumption and, if appropriate, include other factors influencing the individual satisfaction. Likewise, drop-off points are determined in the same way as meeting points, hence in this paper we describe the workflow using only meeting points.

Since the method should be able to serve real-time requests, short response times are crucial. Most of the computationally expensive shortest-path calculations in public transport networks are therefore done in a preprocessing phase. To limit the amount of considered meeting point locations, a GIS workflow is applied. The recommended meeting point is finally selected by a voting rule. In an experimental simulation we demonstrate our algorithm and compare different configuration settings.

\FloatBarrier
\section{Proposed method}
\label{sec:Method}
This section describes the workflow of the proposed method. It is mainly divided into three parts:

\begin{enumerate}
\item \textbf{Preparation phase}\label{step:preparation}
\item \textbf{Precomputation phase}\label{step:precomputation}
\item \textbf{Operation phase}\label{step:operation}
\end{enumerate}

\begin{figure}
    \begin{center}
    \includegraphics[scale=1.5]{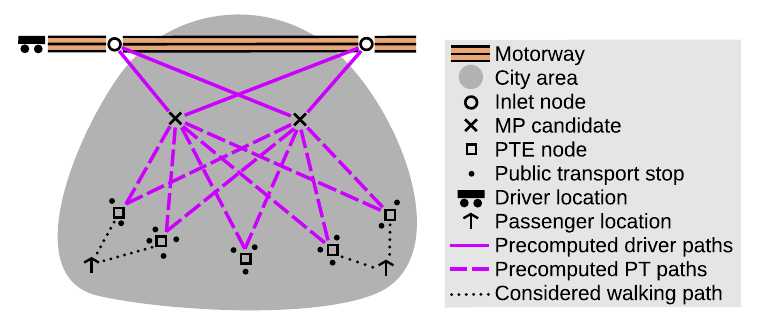}
    \caption{Schematic visualization of the precomputed paths}
    \label{fig:Schemazeichnung2}
    \end{center}
\end{figure}

In a nutshell, the algorithm works as follows. When a request arrives, time costs are iteratively determined for every feasible meeting point candidate in the city. To speed up the checking procedure, the travel times for the driver (driving with vehicle) and for the passengers (walking and using public transport) are precomputed and stored in a matrix. The driving times are stored from \emph{inlet points} on the motorway to every meeting point candidate. On the passenger side, it is computationally inapplicable to precompute the travel times from all possible passenger origins. Hence, representative \textit{public transport entry} (PTE) nodes $\pi \in \Pi$ are created. A request from a passenger then requires first a reachability analysis of PTE nodes in the vicinity of the current location. Subsequently, the precomputed public transport connections from these PTE nodes can be used to estimate the arrival time at the meeting point candidates. Figure \ref{fig:Schemazeichnung2} visualizes the basic principle. The three steps are explained in detail below.

%%%%%%%%%%%%%%%
% PREPARATION %
%%%%%%%%%%%%%%%
\FloatBarrier
\subsection{Preparation phase}
\label{subsec:Preparation}
\begin{figure}
    \begin{center}
    \includegraphics[scale=.50]{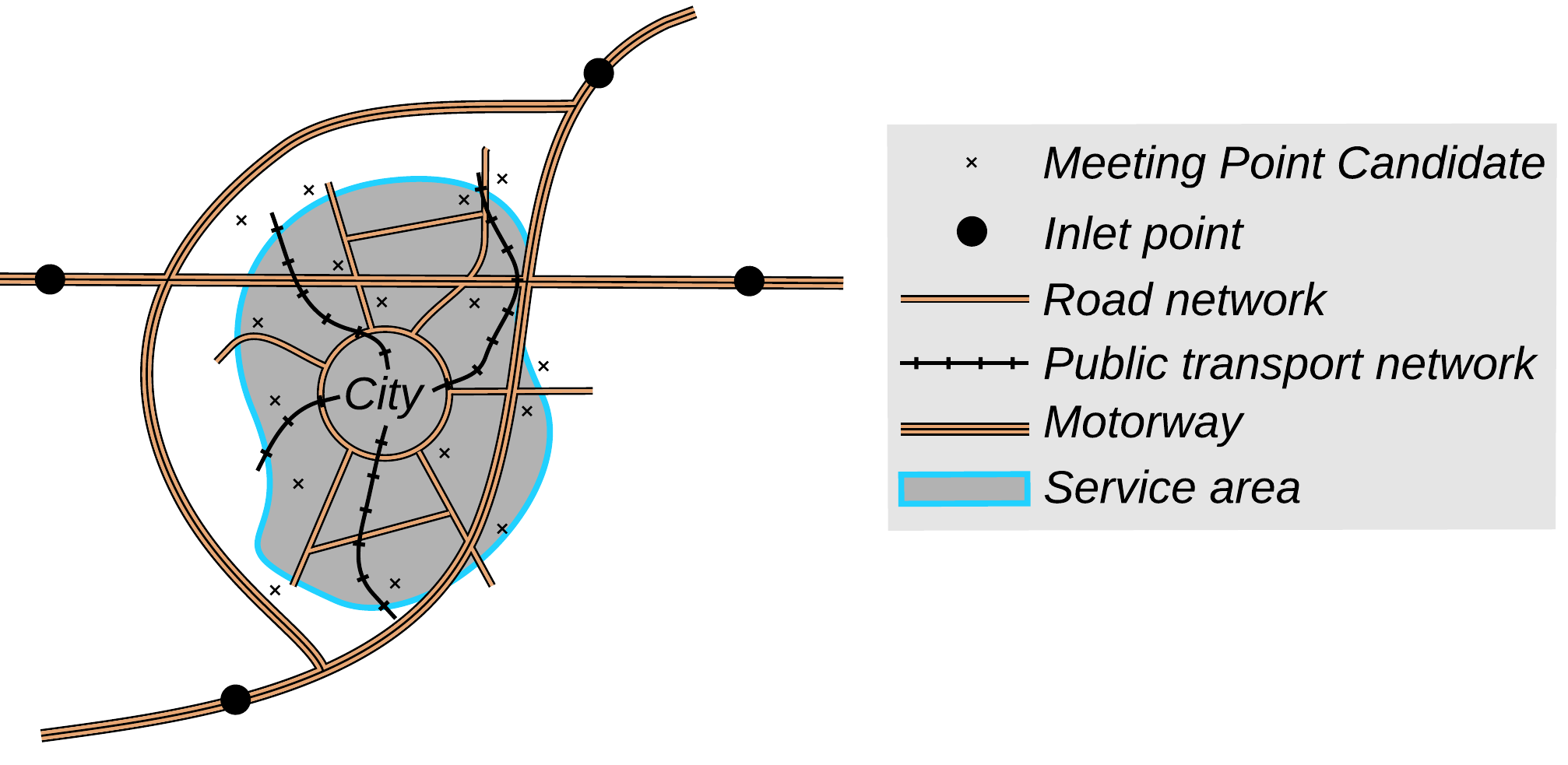}
    \caption{Schematic representation of the service area and location of inlet points}
    \label{fig:Schemazeichnung}
    \end{center}
\end{figure}

In the preparation phase the raw data is processed to prepare the precomputation. The following data is necessary:

\begin{itemize}
\item A street network $G = (V, E)$ of the service area with a set of $V$ vertices and a set of $E$ edges. Each edge $(v^{+},v^{-}) \in E$ has an associated non-negative length $d(v^{+},v^{-})$ and further informations like speed limits, enabling to estimate vehicle driving times $t^{driv}(v^{+},v^{-})$ and passenger walking times $t^{walk}(v^{+},v^{-})$.
\item Public transport (PT) timetable data; typically modeled as time-expanded or time-dependent graph \citep{pyrga2008efficient, bast2015route}. Following the notation of \cite{muller2007timetable}, we model the timetable data as a set of stops $\mathcal{S}$, a set of vehicle lines $\mathcal{Z}$ (e.g. tram line) and a set of elementary connections $\mathcal{C}$. A connection element $c \in \mathcal{C}$ is then a 5-tuple $c = (z,s^{+},s^{-},t^{+},t^{-})$ that can be interpreted as vehicle $z$ leaving stop $s^{+}$ at time $t^{+}$ and arriving stop $s^{-}$ at time $t^{-}$.
\item A set of meeting point candidate locations $M$ in the service area.
\item A set of \textit{Inlet points} $I$. These points should be located on motorways or other high-level roads and mark the entry (and exit, respectively) of the service area such that the  important interurban connections pass an inlet point inbound and another inlet point outbound (see Figure \ref{fig:Schemazeichnung}).
\end{itemize}

The preparation phase can further be subdivided into the preparation of the driving network (step \ref{subsubsec:Preparation_G}), the generation of public transport entry (PTE) nodes (step \ref{subsubsec:Preparation_PTE}) and the preparation of meeting point candidates (step \ref{subsubsec:Preparation_MPC}).

\subsubsection{Street network preparation}
\label{subsubsec:Preparation_G}
The street network itself does not require much preparation, besides the connection of meeting point candidate nodes $M$ and inlet nodes $I$ to the graph $G$. However, in real-world street networks, the travel times are often time-dependent, meaning that the arrival time to an edge determines the actual time to traverse the edge \citep{demiryurek2010case}. Streets are often congested during rush hours, so the travel times are then significantly higher than during off-peak times. The usage of time-dependent shortest-path (TDSP) models is hence recommended but not required for the proposed method. In the last decades, numerous methods have been proposed to model TDSP, e.g. as discrete-time algorithms \citep{cai1997time, chabini1998discrete} or as continuous-time algorithms \citep{orda1990shortest}. 

\FloatBarrier
\subsubsection{Generation of public transport entry (PTE) nodes}
\label{subsubsec:Preparation_PTE}
In the precomputation step, the public transport connections from PTE nodes to meeting point candidates will be determined. Since the passengers are supposed to use the public transportation system and the stops are where they change over from walking to public transport, it is reasonable to place the PTE nodes close to the public transport stops $\mathcal{S}$. A trivial way is to simply create a PTE node $\pi$ at every stop $s$. However, since every PTE node invokes a precomputation and storage of connections in step \ref{step:precomputation}, it is useful to reduce the amount of PTE nodes beforehand.

\begin{figure}
  \begin{subfigure}[b]{0.5\linewidth}
    \centering
    \includegraphics[width=0.9\linewidth]{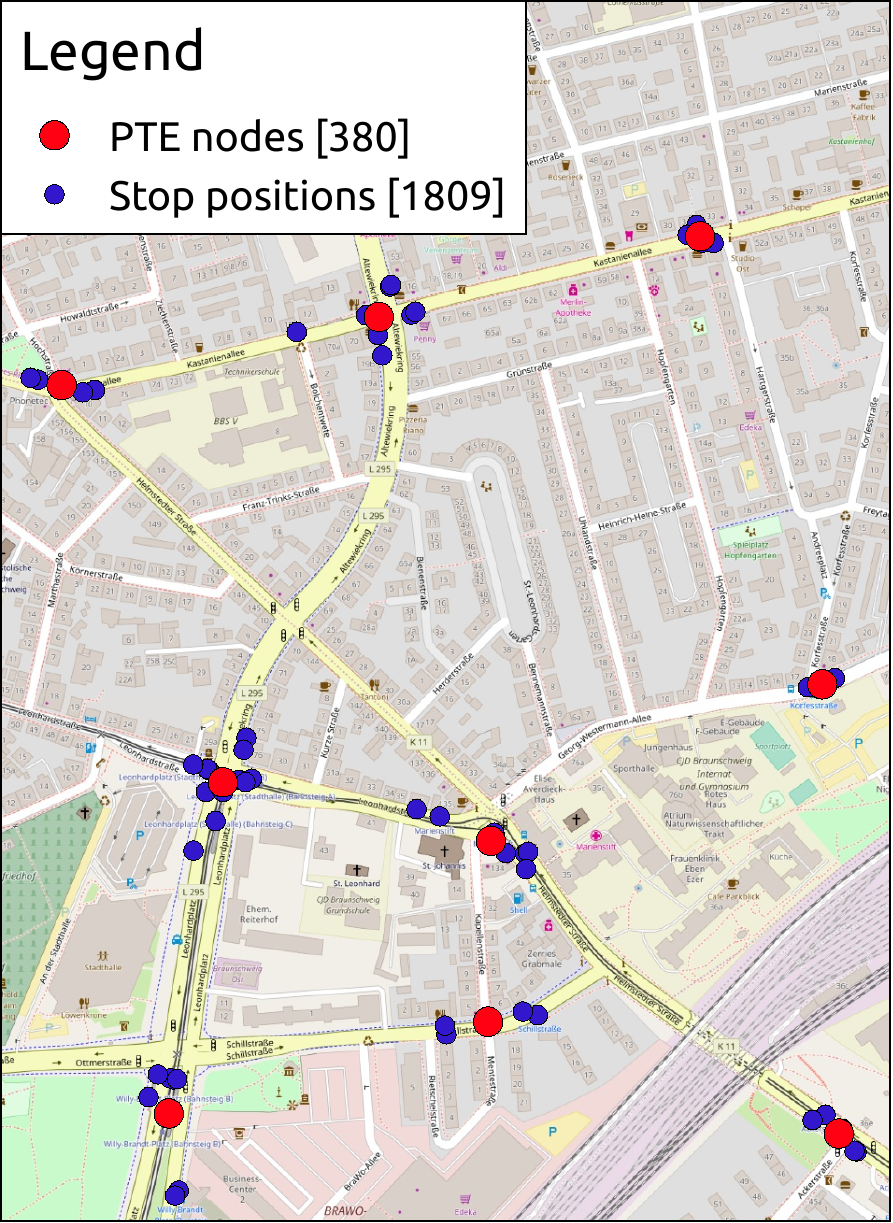} 
    \caption{Clustering of stop positions.} 
    \label{fig:preparation1} 
    \vspace{1ex}
  \end{subfigure}
  \begin{subfigure}[b]{0.5\linewidth}
    \centering
    \includegraphics[width=0.9\linewidth]{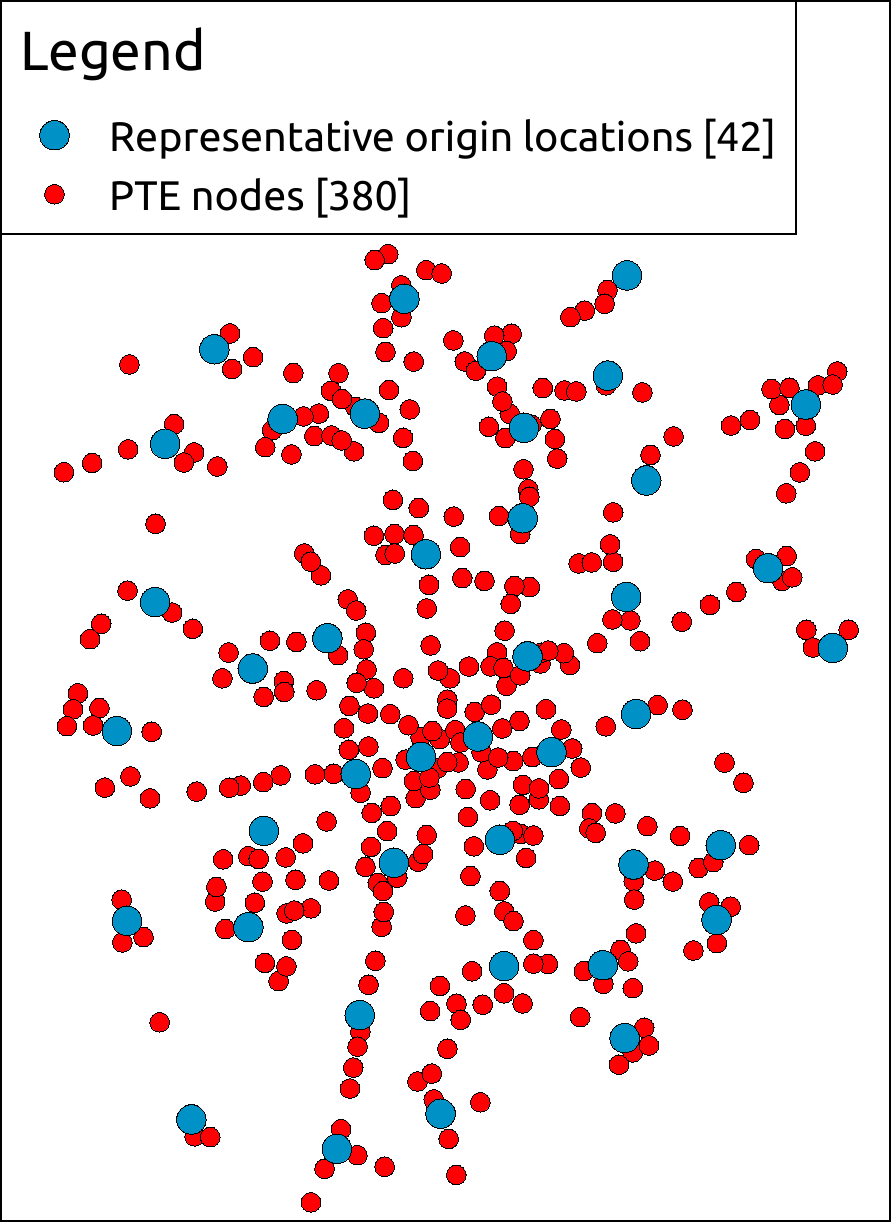} 
    \caption{Sampling among stop positions.} 
    \label{fig:preparation2} 
    \vspace{1ex}
  \end{subfigure}%%
  \caption{Stop preparation.}
  \label{fig:stop_preparation} 
\end{figure}

To this end, we propose to group the stop positions $\mathcal{S}$ first. This is advisable since often one stop (e.g. ``Main Station'') consists of several discrete stopping positions, e.g. for bus, tram and different directions. The grouping should integrate these stopping places, either based on the stop name, or, if the naming is not consistent, by a density-based clustering like DBSCAN~\citep{ester1996density}. Figure \ref{fig:preparation1} shows the result of such a clustering. The DBSCAN distance threshold $\epsilon$ should be chosen such that each group of stopping positions covers exactly one stop. In our simulation experiments (section \ref{sec:Simulation}), 100 m was determined as appropriate value for this purpose. However, if stop positions of two different stops are closer than $\epsilon$, the DBSCAN clustering will group them togehter. Hence, a postprocessing check is recommended to ensure that all stop positions are covered by a PTE node $\pi \in \Pi$ within a reasonable distance.

\subsubsection{Meeting point candidates preparation}
\label{subsubsec:Preparation_MPC}
For the MP preparation phase we assume the meeting point candidates $M$ to be already predetermined, e.g. all parking places or petrol stations of a city.

Analogue to the previous step \ref{subsubsec:Preparation_PTE}, it is possible to carefully reduce the amount of MP candidates by an order of magnitude to a reasonable size. Since many of the initially identified candidate locations will be located at inappropriate locations and are hence not very useful, it is advisable to filter them before the precomputing phase. 

\begin{figure}
  \begin{subfigure}[b]{0.5\linewidth}
    \centering
    \includegraphics[width=0.9\linewidth]{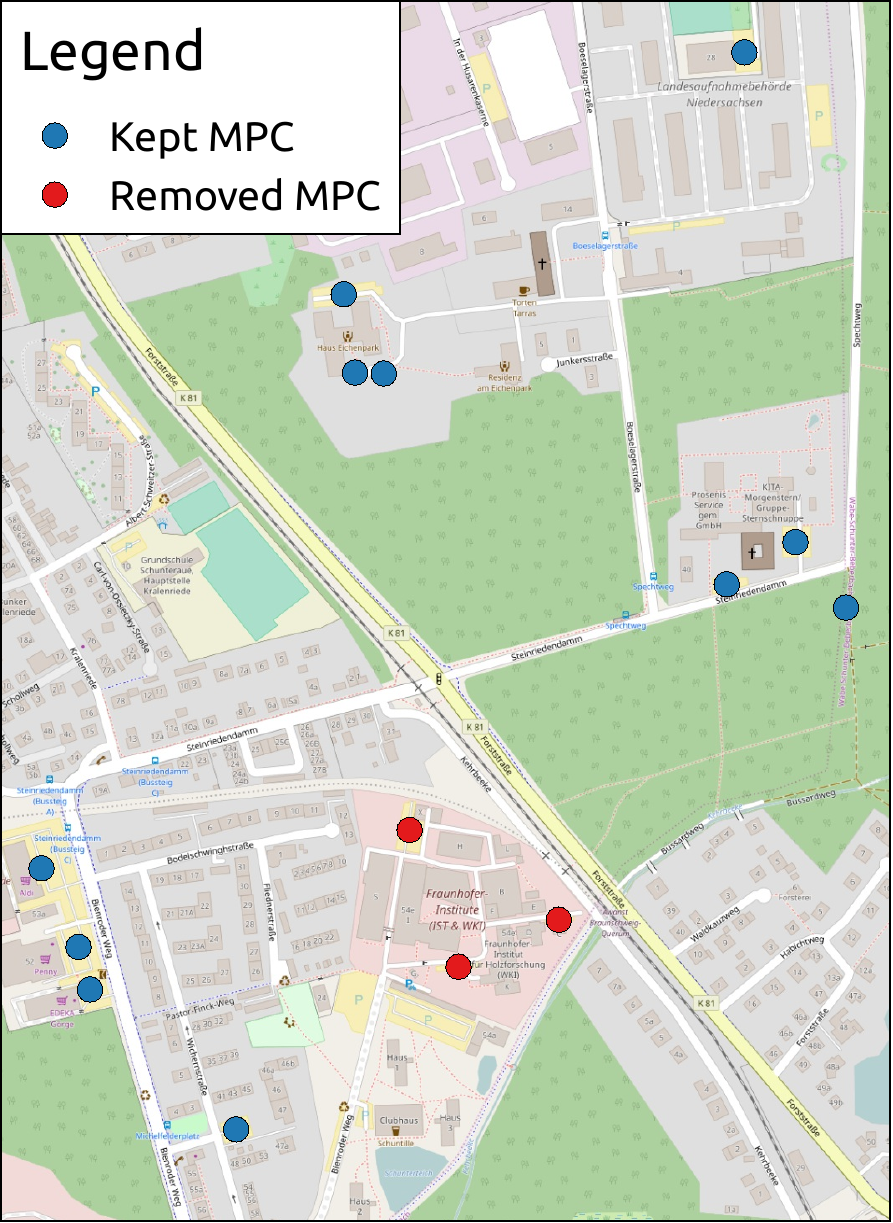} 
    \caption{Filtering of unreachable meeting points, e.g. due to private property} 
    \label{fig:preparation3} 
    \vspace{1ex}
  \end{subfigure}
  \begin{subfigure}[b]{0.5\linewidth}
    \centering
    \includegraphics[width=0.9\linewidth]{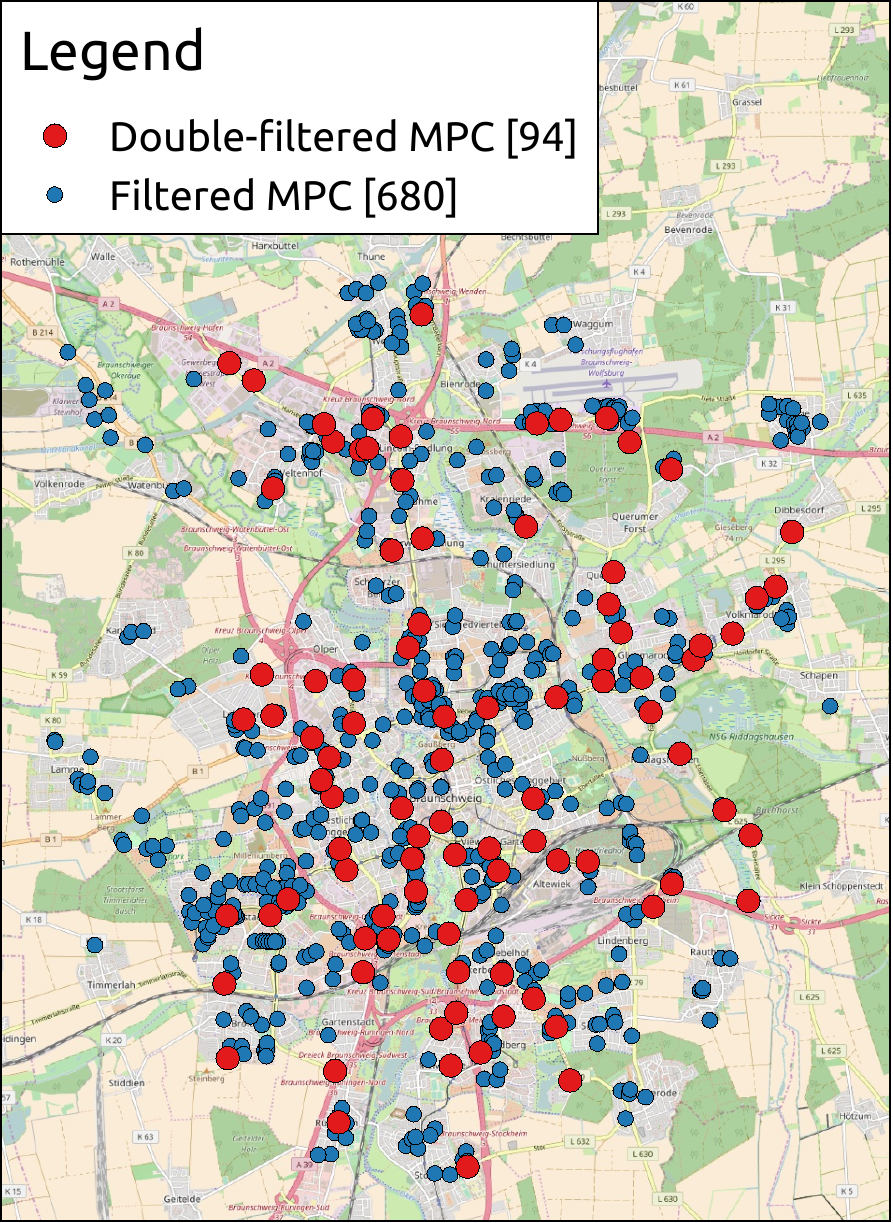} 
    \caption{Filtering of unused meeting points after a simulation run} 
    \label{fig:preparation4} 
    \vspace{1ex}
  \end{subfigure}%%
  \caption{Meeting point candidates preparation}
  \label{fig:mpc_preparation} 
\end{figure}

For this, firstly all meeting point candidates that have no stop position $s \in \mathcal{S}$ within a reasonable walking threshold $d^{walk}_{*}$ are removed. Secondly, all candidates that are unreachable by drivers or passengers due to access restrictions are removed, e.g. if they are located on private property (see Figure \ref{fig:preparation3}). If the municipal traffic management wants to keep the ride-sharing traffic off the city center or other areas, this is also the right step to manually remove further undesirable locations.

All points of the remaining set are in theory feasible for being considered; however, a majority of these points is still not useful to keep since they are unlikely to be ever used in the operational phase. Hence, we propose a simulation run based on evenly sampled passenger origins to determine the usage frequency of the meeting points and keep only the most promising. A few representative passenger origin locations $\lambda \in \Lambda$ need to be sampled as starting points. This step can be done manually or automatized. In our simulation (section \ref{sec:Simulation}) we use an iterative k-Means clustering of the PTE nodes $\Pi$ for this, as outlined in Algorithm 1 in appendix \ref{algs}. Figure \ref{fig:preparation2} shows the result of the k-Means clustering.

The resulting location set $\Lambda$ represents possible passenger origins all over the service area (e.g. such that at least one location is in every suburb). Then, the travel times to the meeting point set $M$ are precomputed (see section \ref{subsec:Phase2}). Subsequently, fictive meetings of random driver/passenger groups in the service area are simulated and the recommended meeting points recorded. More detailed, a list of $n$ tuples, each containing a random driver inbound inlet node $i^{+}$, a random driver outbound inlet node $i^{-}$, a random group of representative passenger entry nodes $\Lambda$, and a random time of day, is created and iteratively used as simulation input. 

\begin{figure}
    \begin{center}
    \includegraphics[scale=.7]{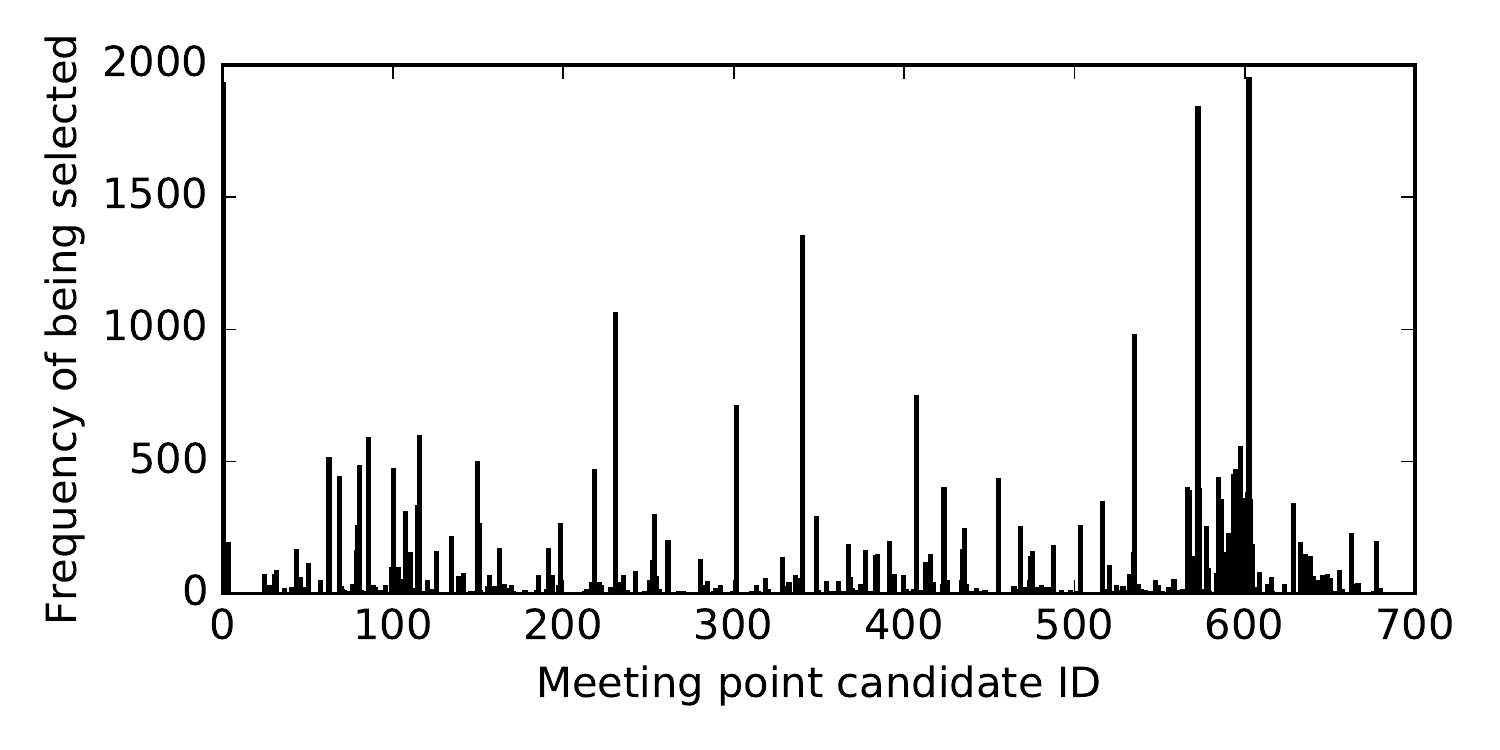}
    \caption{Frequency of meeting point candidates being selected (Total runs: 36000)}
    \label{fig:MPC_Usage}
    \end{center}
\end{figure}

Figure \ref{fig:MPC_Usage} shows a typical frequency of meeting points being selected in the simulation. As can be seen, the meeting point candidates are chosen with a very different frequency - some very often, others never. The meeting points with low scores are then removed by a threshold selection, yielding a reduced set $M' \subset M$.

However, the remaining meeting points $\mu \in M'$ can still be located very close to each other. In these situations, only one of the adjacent meeting points would be sufficient. Hence, the set of candidate points can further be downsized. A refined approach based on the one previously described includes an initial spatial clustering of the meeting point candidates by DBSCAN. In the subsequent analysis of the selection frequency, only the meeting point having the highest score among its direct neighbours is kept. The proposed method is outlined in Algorithm 2 in appendix \ref{algs}. Figure \ref{fig:preparation4} illustrates the final reduction of meeting point candidates.

%%%%%%%%%%%%%%%%%%
% PRECOMPUTATION %
%%%%%%%%%%%%%%%%%%
\FloatBarrier
\subsection{Precomputing phase}
\label{subsec:Phase2}

In this phase, travel times are precomputed and stored in different matrices. 

Firstly, the driving times $t^{driv}$ from all inlet points $I$ to all meeting point candidates $M'$ and back are stored in matrices $A_{\Psi}$. Equation~\ref{eq:driverMatrix} shows the matrix structure for static driving times that do not change over time.

Secondly, a matrix $A_{P}$ is created, containing all possible multimodal public transport connections throughout the day from all PTE nodes in $\Pi$ to all meeting point candidates in $M'$. In this matrix, each entry is a list of departure times $t^{+}$ and arrival times $t^{-}$, sorted by departure time (see Equation~\ref{eq:passengerMatrix}). Slow connections that are being overtaken by other connections are removed, so that only non-overlapping connections are stored. If it is possible to walk from a PTE node $\pi$ to a MPC $\mu$, only the walking time is stored.

\begin{equation}
A^{inbound}_{\Psi} = \bordermatrix{
  & \mu_{1} & \mu_{2} \cr
i^{+}_{1} & t^{driv} & t^{driv} \cr
i^{+}_{2} & t^{driv} & t^{driv} \cr},
\qquad
A^{outbound}_{\Psi} = \bordermatrix{
  & i^{-}_{1} & i^{-}_{2} \cr
\mu_{1} & t^{driv} & t^{driv} \cr
\mu_{2} & t^{driv} & t^{driv} \cr}
\label{eq:driverMatrix}
\end{equation}

\begin{equation}
A_{P} = \bordermatrix{
  & \mu_{1} & \mu_{2} \cr
\pi_{1} & \left[\left(t^{+}_{1}, t^{-}_{1}\right), \left(t^{+}_{2}, t^{-}_{2}\right), \dots\right] & \left[\left(t^{+}_{1}, t^{-}_{1}\right), \dots\right] \cr
\pi_{2} & \left[\left(t^{+}_{1}, t^{-}_{1}\right), \left(t^{+}_{2}, t^{-}_{2}\right), \dots\right] & t^{walk} \cr}
\label{eq:passengerMatrix}
\end{equation}

\subsection{Operational phase}
\label{subsec:Phase3}
The operational module can be regarded as a service interface waiting for incoming requests of a driver/passenger group that returns a recommendation of one (or more) meeting points. The necessary components of a request are:
\begin{itemize}
\item Planned driver inlet node (inbound) $i^{+}$
\item Planned driver inlet node (outbound) $i^{-}$
\item Current location of the driver $\lambda\left(\psi\right)$
\item Current (or planned) cocation of one or multiple passengers $\lambda\left(\rho\right)$
\item Maximum driver detour time $t^{detr}_{*}$
\item Waiting time tolerance $t^{wait}_{*}$
\item Maximum passenger walking distance $d^{walk}_{*}$
\end{itemize}

The maximum driver detour parameter $t^{detr}_{*}$ controls the feasible meeting point candidates, i.e. a smaller value of $t^{detr}_{*}$ leads to selection of meeting points that are closer to the motorway exits. The parameter allows drivers to specify their time budget for picking up (or dropping off) passengers. Furthermore, the parameter can be used by traffic management entities to influence how far vehicles should penetrate the city, e.g. for pollution reduction.

The waiting time tolerance parameter $t^{wait}_{*}$ defines the flexibility of arrival times at the meeting point. A negative value of -5 minutes indicates that all passengers must arrive at the meeting point at least 5 minutes prior to the driver arrival. In contrast, a positive value of 5 minutes allows the passengers to arrive up to 5 minutes later than the driver.

The workflow of request processing contains the following steps:
\begin{enumerate}
\item Estimate driver arrival times at meeting point candidates
\item Determine reachable PTE nodes for the passengers
\item Estimate passenger arrival times at meeting point candidates
\item Compute total travel times
\item Voting
\end{enumerate}

\subsubsection{Estimate driver arrival times at meeting point candidates}
\label{subsubsec:DriverArrivalTimes}
In order to calculate all arrival and departure times, the only unknown value is the expected time of the driver passing the inlet node (inbound) $t^{-}_{\psi}(i^{+})$. It can be estimated based on the current location of the driver $\lambda(\psi)$, which can automatically be transmitted from any GPS sensor. An arbitrary (third-party) routing service is further applied to estimate the remaining journey time until the inlet node is reached. Using Equation \ref{eq:driverArrivalAtMP} and the precomputed driving time matrix $A^{inbound}_{\Psi}$, the arrival times of the driver at all meeting point candidates can then instantly be estimated:
\begin{equation}
t^{-}_{\psi}(\mu) = t^{-}_{\psi}(i^{+}) + A^{inbound}_{\Psi}(i^{+} \rightarrow \mu)
\label{eq:driverArrivalAtMP}
\end{equation}
If the threshold $t^{detr}_{*}$ is set, all meeting point candidates that require a driver detour time exceeding the threshold can be disregarded for this request.

\subsubsection{Determine reachable stop nodes for the passengers}
\label{subsubsec:DetStopNodes}
Since the algorithm is designed as a location-based service, the selected meeting point depends on the current (or planned) position of the passengers $\lambda(\rho) \in P$. The location can, analogue to the driver, automatically be transmitted from any GPS sensor, e.g. a smartphone. In addition, also a manual location input of the customer is possible, which may be useful if the location at time of departure is already known beforehand. Since the public transport connections are only precomputed from the PTE nodes $\Pi$, as described in Section \ref{subsec:Phase2}, first all reachable PTE nodes $\Pi_{\rho}$ have to be determined with respect to the maximum walking distance $d^{walk}_{*}$. Further, for each PTE node $\pi \in \Pi_{\rho}$, the corresponding walking time $t^{walk}$ has to be calculated.

If, due to privacy reasons, the actual location of the passenger should be obfuscated, this step can also be outsourced to a trusted third-party that returns the closest PTE nodes, or the passengers choose the PTE nodes directly and estimate the walking time by themselves.

\subsubsection{Estimate passenger arrival times at meeting point candidates}
\label{subsubsec:PassengerArrivalTimes}
The previously derived driver arrival times at meeting points dictate the passenger connections to be considered. For every reachable PTE node of the passengers and every meeting point, the corresponding time of departure for the passenger is calculated such that the waiting time constraint is met. For this, the item $\left(t^{+}(\pi), t^{-}(\mu)\right)$ closest to the arrival time is fetched from the sorted list of arrival times in the passenger connection matrix $A^{P}$, which can be efficiently done using binary search. Note that the waiting time tolerance $t^{wait}_{*}$ has to be applied. Consequently, the departure time for this passenger can be determined by Equation~\ref{eq:passengerDepartureAtLocation}, including the initial walking time from the current location to PTE node.
\begin{equation}
t^{+}_{\rho}\left(\lambda\left(\rho\right)\right) = t^{+}(\pi) - t^{walk}\left(\lambda\left(\rho\right) \rightarrow \pi\right)
\label{eq:passengerDepartureAtLocation}
\end{equation}

\subsubsection{Compute total travel times for all meeting point candidates}
\label{subsubsec:TotalTravelTimes}
Having the travel times for the driver and all passengers at hand, the total travel time can be computed for each feasible meeting point candidate by summing up the values (Equation~\ref{eq:totalTravelTime}). The total travel time includes the time to reach the meeting point as well as the driving time from the meeting point to the outbound node (Equation~\ref{eq:driverArrivalAtOutboundNode}).
\begin{equation}
  t^{-}_{\psi}(i^{-}) = \max \left( t^{-}_{\psi}\left(\mu\right), \max \left( \{t^{-}_{\rho}\left(\mu\right) \mid \rho \in P \} \right) \right) + A^{outbound}_{\Psi}(\mu \rightarrow i^{-})
\label{eq:driverArrivalAtOutboundNode}
\end{equation}
\begin{equation}
t^{total} = t^{-}_{\psi}\left(i^{-}\right) - t^{+}_{\psi}\left(i^{+}\right) + \sum_{\rho \in P} \left(t^{-}_{\psi}\left(i^{-}\right) - t^{+}_{\rho}\left(\lambda\left(\rho\right)\right)\right)
\label{eq:totalTravelTime}
\end{equation}

\subsubsection{Voting}
\label{subsubsec:Voting}
The last step of the workflow is to choose an appropriate meeting point among the set of candidates. Considering several persons who need to agree on a meeting point, those persons will probably have different preferences regarding the possible meeting points, not just because of their different distances to the meeting points but also because of other properties such as (subjective) safety at the meeting points, prominence, sheltering possibilities, accessibility etc. \citep{Czioska2017314}.

A straightforward approach for reaching a socially acceptable agreement based on differing individual possibilities is the application of voting rules, i.e. to reach a common decision, an election is held. In the literature, different voting rules for decision making in traffic applications can be considered \citep{dennisen2015agent}. 

Formally, a voting rule with exactly one winner can be defined as follows \citep{rothe2012einfuhrung}: An \emph{election} or \emph{preference profile} is a tuple $(C,V)$ with $C$ set of candidates and $V$ list of voters, where each voter is represented via their vote which specifies their preferences regarding the candidates in $C$. A voting rule with exactly one winner is a \emph{social choice function} 
\begin{equation}
f: \{(C,V) \mid (C,V) \textrm{ is a preference profile}\} \rightarrow C
\end{equation}
which assigns to each given preference profile exactly one winner.

There are several possibilities for transforming individual preferences into votes for the election. For example, we could ask each person to specify a complete ranking over the possible meeting points based on their preferences. However, in our simulation we disregard aspects such as safety, prominence etc. due to simplicity, and derive the votes directly from the travel times of the respective person to the meeting points.

We propose to use two voting rules. The first is a \emph{range voting} rule, where each voter scores all candidates on a range ballot based on the inverse travel time. The scores are summed up, and the candidate with the highest value is elected, i.e. the result is the meeting point having the lowest value of $t^{total}$. This corresponds to optimizing the travel times according to an utilitarian approach.

However, this rule does not necessarily yield the most socially acceptable solution. Consider a situation with three riders A, B and C and two meeting point candidates $\alpha$ and $\beta$. We can interpret the travel times as dissatisfaction values. Meeting point $\alpha$ gets a dissatisfaction score of 10 by each rider, and meeting point $\beta$ is scored with a dissatisfaction of 2 by riders A and B. Rider C however scores meeting point $\beta$ with a dissatisfaction of 25. According to the range voting rule, $\beta$ wins because $t^{total}(\beta) = 2 + 2 + 25 \leq t^{total}(\alpha) = 10 + 10 + 10$. Now it can be argued that the selection discriminates rider C. The example hence highlights that the range voting rule is prone to imbalanced travel times among the riders.

For this reason we propose a second election principle following the \emph{minimax} principle. Here, the meeting point that minimizes the maximum travel time (or detour time of the driver, respectively) among the riders is chosen, leading to a more balanced distribution of travel times. The travel times are regarded as dissatisfaction values, and the winner of the election is a meeting point with the lowest maximum dissatisfaction value. This corresponds to optimizing the travel time according to an egalitarian approach. In our basic example, meeting point $\alpha$ would then have been recommended because $10 \leq 25$. In the simulation experiment (Section~\ref{subsec:Experiment_Results}), we compare the results of these two methods.

\FloatBarrier
\section{Simulation}
\label{sec:Simulation}
In order to demonstrate the effect of our algorithm we conducted a simulation experiment using a set of randomly generated trip requests.

\subsection{Simulation setting and data}
\label{subsec:simulation_setting}
For our simulation we use the medium-sized city of Braunschweig ($\sim$ 250.000 inhabitants) as a model. The city centre is dominated by the historical core and a pedestrian precinct. It is surrounded by a ring road and some densely populated areas. In the outskirts, the population density is significantly lower. In addition, there are some industrial areas. For vehicles, there is an outer ring formed by five motorways, with no motorway on the eastern side of the city. The public transport system of Braunschweig\footnote{\url{http://www.verkehr-bs.de/}} includes a tram network with 5 tram lines and a bus network with 37 bus lines. The main public transport lines are in operation all day long with a night break from approx. 2am to approx. 4am.

\subsubsection{Travel times}
All travel times for driving, walking and public transport have been computed with an instance of OpenTripPlanner\footnote{\url{http://www.opentripplanner.org/}}, a JAVA-based open-source multimodal routing engine. The necessary data for the street network was obtained from OpenStreetMap\footnote{\url{http://www.openstreetmap.org/}}. The timetable information and the stop locations were obtained in GTFS format from an open data pool provided by Connect GmbH\footnote{\url{http://www.connect-fahrplanauskunft.de}}. After the preparation phase (see section \ref{subsubsec:Preparation_PTE}), 380 PTE nodes are remaining.

\subsubsection{Meeting Points}
As meeting point candidates we extracted all petrol stations and the centroids of parking places without fees (e.g. in front of restaurants or supermarkets) within the investigation area from OpenStreetMap, since these places offer a safe and convenient boarding possibility. Initially, 705 meeting point candidates have been extracted.
After all refinement steps (section \ref{subsubsec:Preparation_MPC}), 94 meeting point candidates remain (see Figure \ref{fig:BS_Map}). As can be seen, the meeting point candidates in the more remote outskirts are all removed - this is due to the filtering step by usage, since meeting points in these remote areas have been selected too infrequently in the simulation.

\subsubsection{Inlet nodes}
\begin{figure}
\begin{center}
\includegraphics[scale=.6]{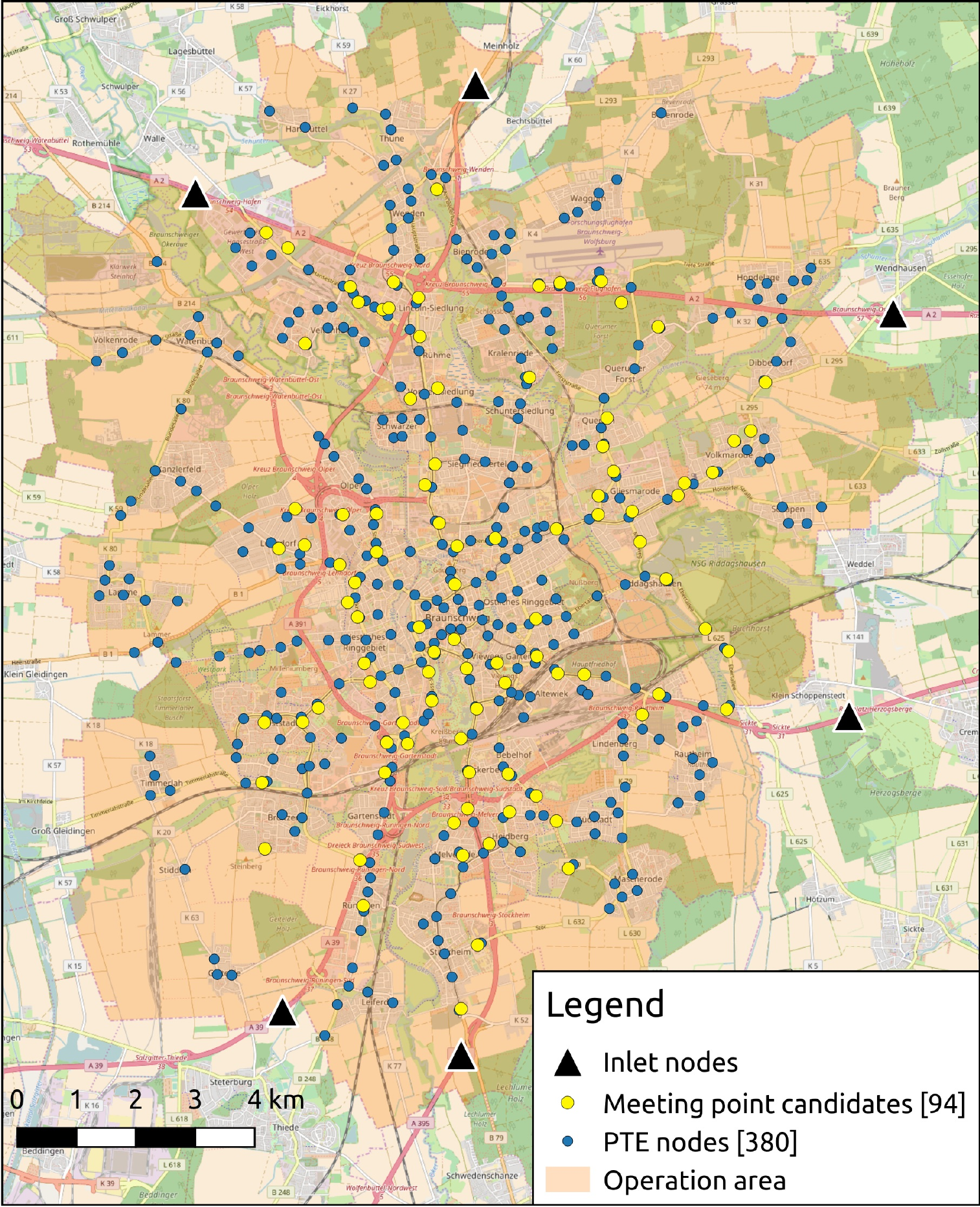}
\caption{Operation area (city of Braunschweig) with inlet nodes, meeting point candidates and PTE nodes. Background: OpenStreetMap}
\label{fig:BS_Map}
\end{center}
\end{figure}

As inbound and outbound nodes we manually selected six locations on the motorways surrounding the city, visualized as black triangles in Figure \ref{fig:BS_Map}.

\subsubsection{Random demand}
\begin{figure}
\begin{center}
\includegraphics[scale=.5]{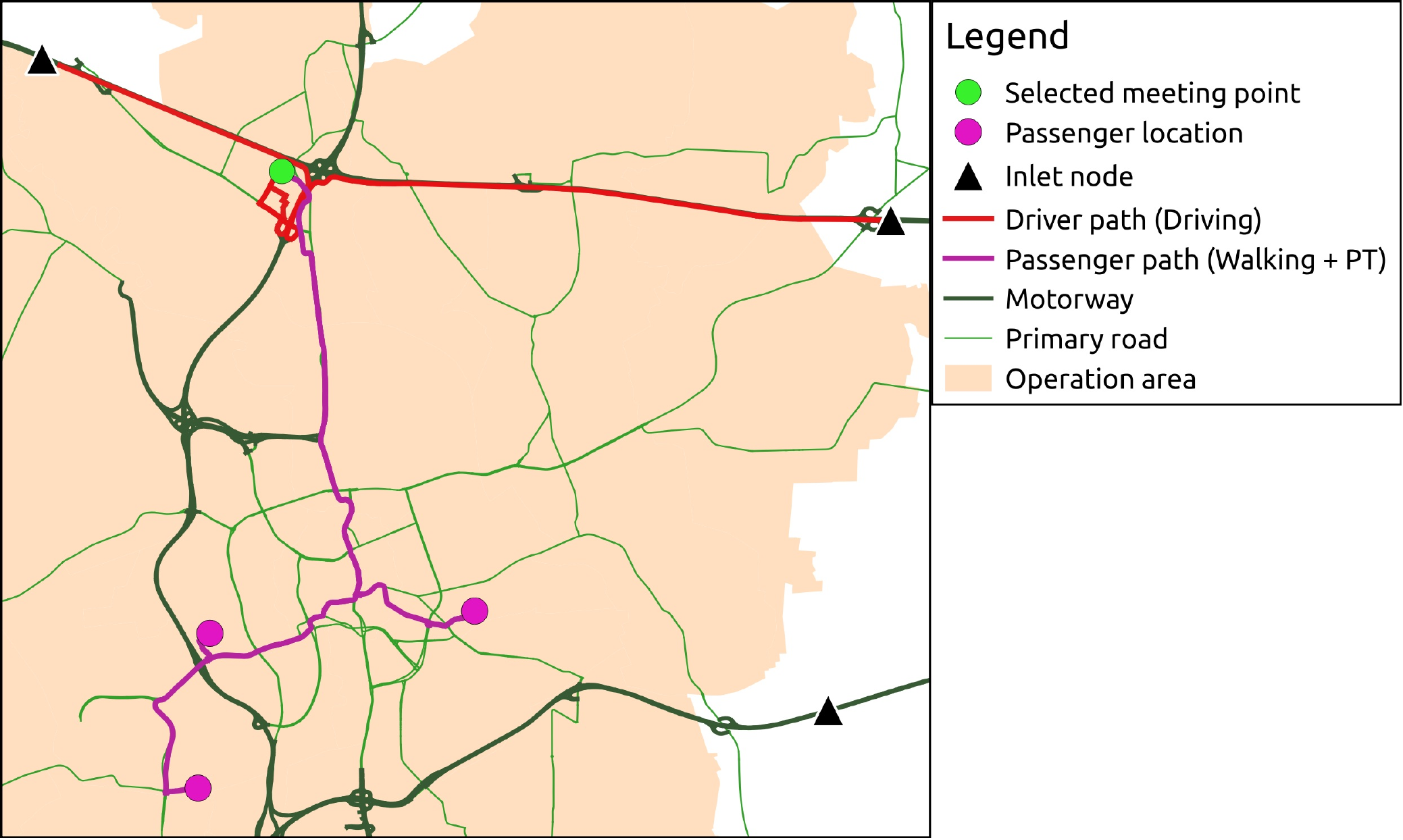}
\caption{Example meeting point selection with three passengers involved}
\label{fig:BS_Paths}
\end{center}
\end{figure}

The driver route is randomly chosen among the six available inlet nodes. U-turns (inbound and outbound inlet nodes are equal) are not allowed. The time of driver arrival at the inbound inlet node is randomly chosen between 6am and 11pm to avoid the night break. The passenger origin locations are randomly sampled based on residential building geometries within the service area. The building information was obtained from the municipality of Braunschweig\footnote{\url{www.braunschweig.de/leben/stadtplanung_bauen/geoinformationen/geoinformationen.html}}. The probability of a building being chosen is dependent on its volume, i.e. bigger buildings are chosen more often than smaller buildings, as it is assumed that more people are living there and thus create a higher demand. A request consists of a driver route and a set of one to three passengers (randomly selected). Figure \ref{fig:BS_Paths} shows an example meeting point recommendation involving three passengers.

\subsection{Results}
\label{subsec:Experiment_Results}
\begin{figure}
  \begin{subfigure}[b]{0.5\linewidth}
    \centering
    \includegraphics[width=0.9\linewidth]{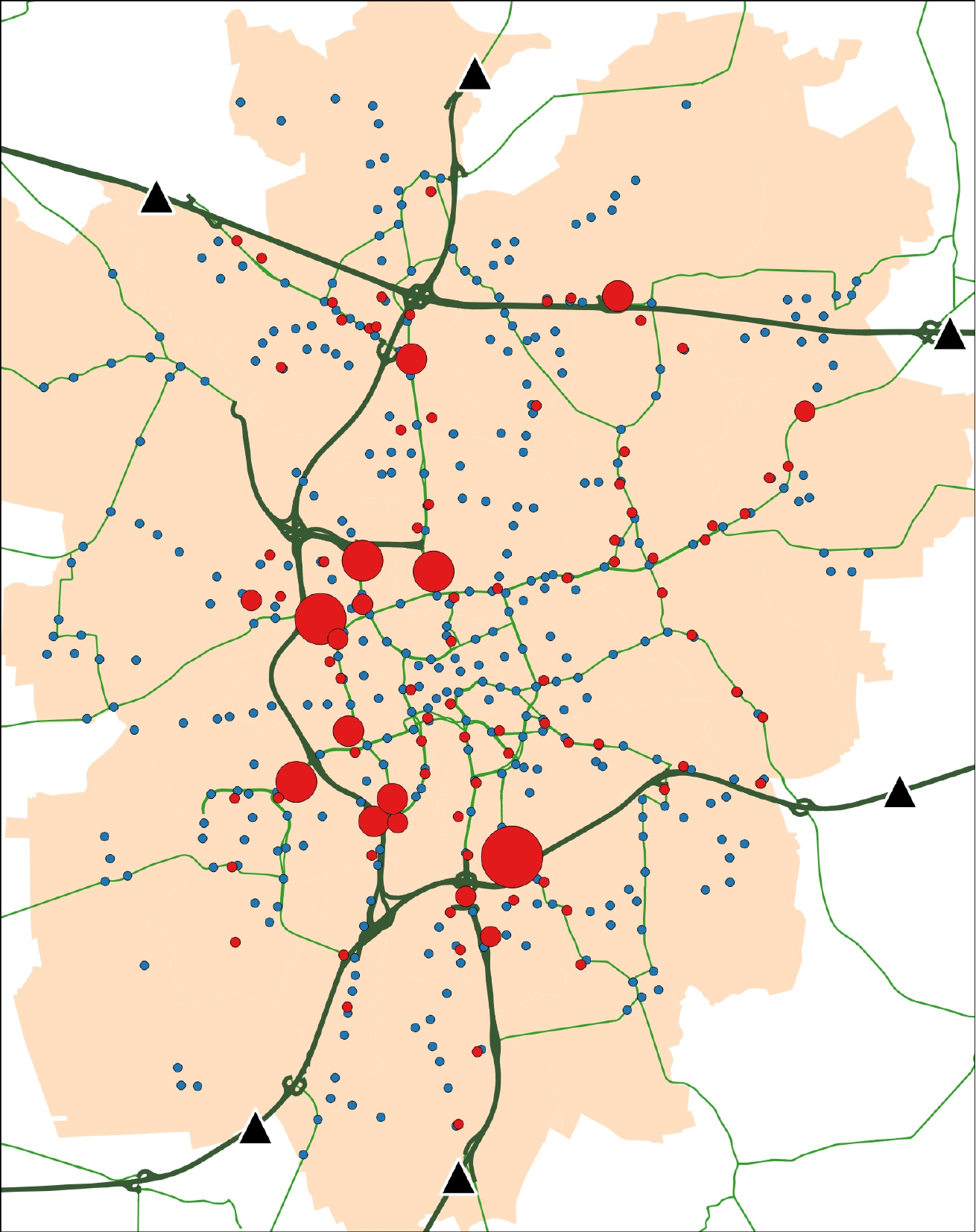} 
    \caption{MP selection frequency (Noon)} 
    \label{subfig:sft1} 
    \vspace{1ex}
  \end{subfigure}
  \begin{subfigure}[b]{0.5\linewidth}
    \centering
    \includegraphics[width=0.9\linewidth]{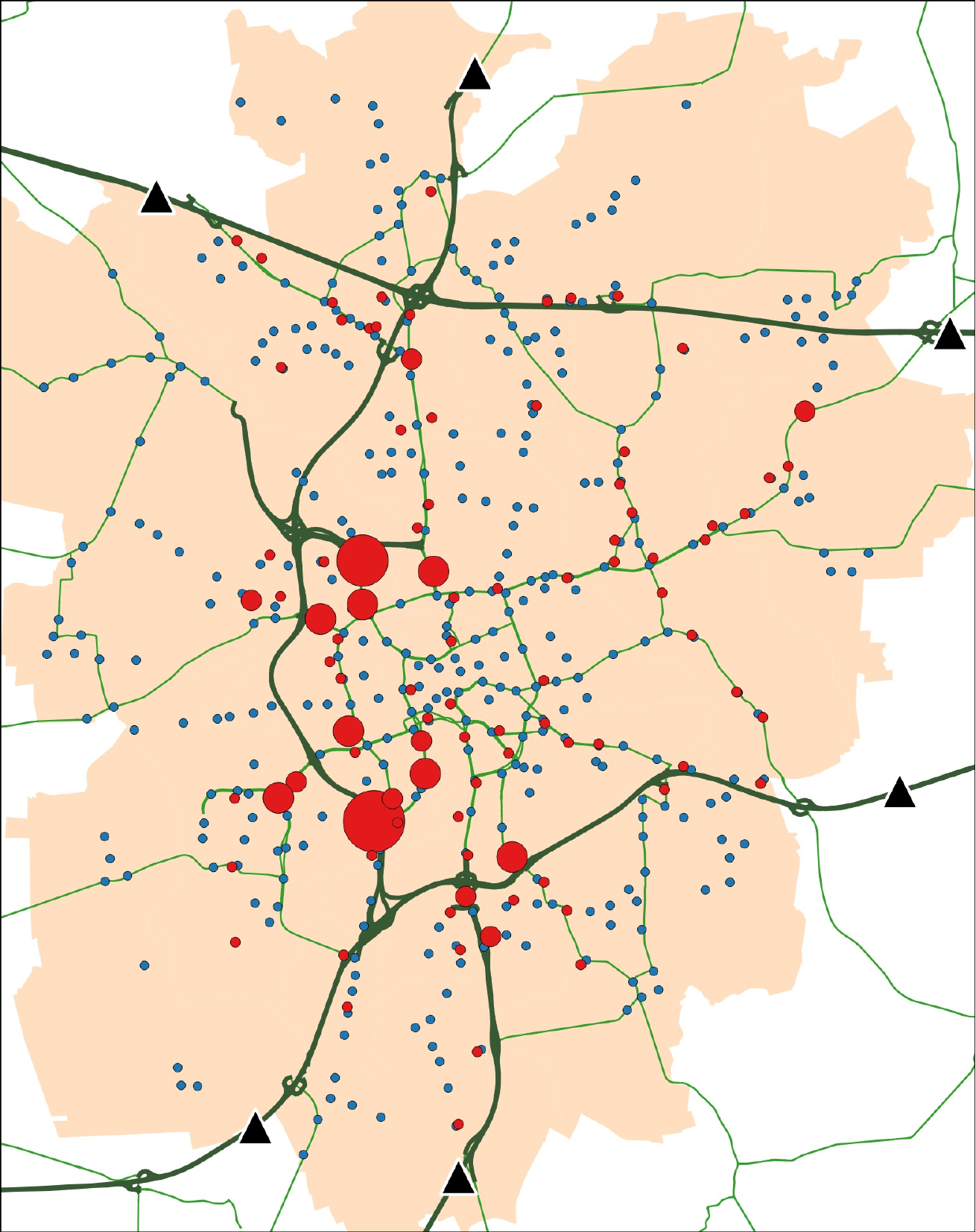} 
    \caption{MP selection frequency (Evening)} 
    \label{subfig:sft2} 
    \vspace{1ex}
  \end{subfigure}
  \newline
  \begin{subfigure}[b]{0.5\linewidth}
    \centering
    \includegraphics[width=0.9\linewidth]{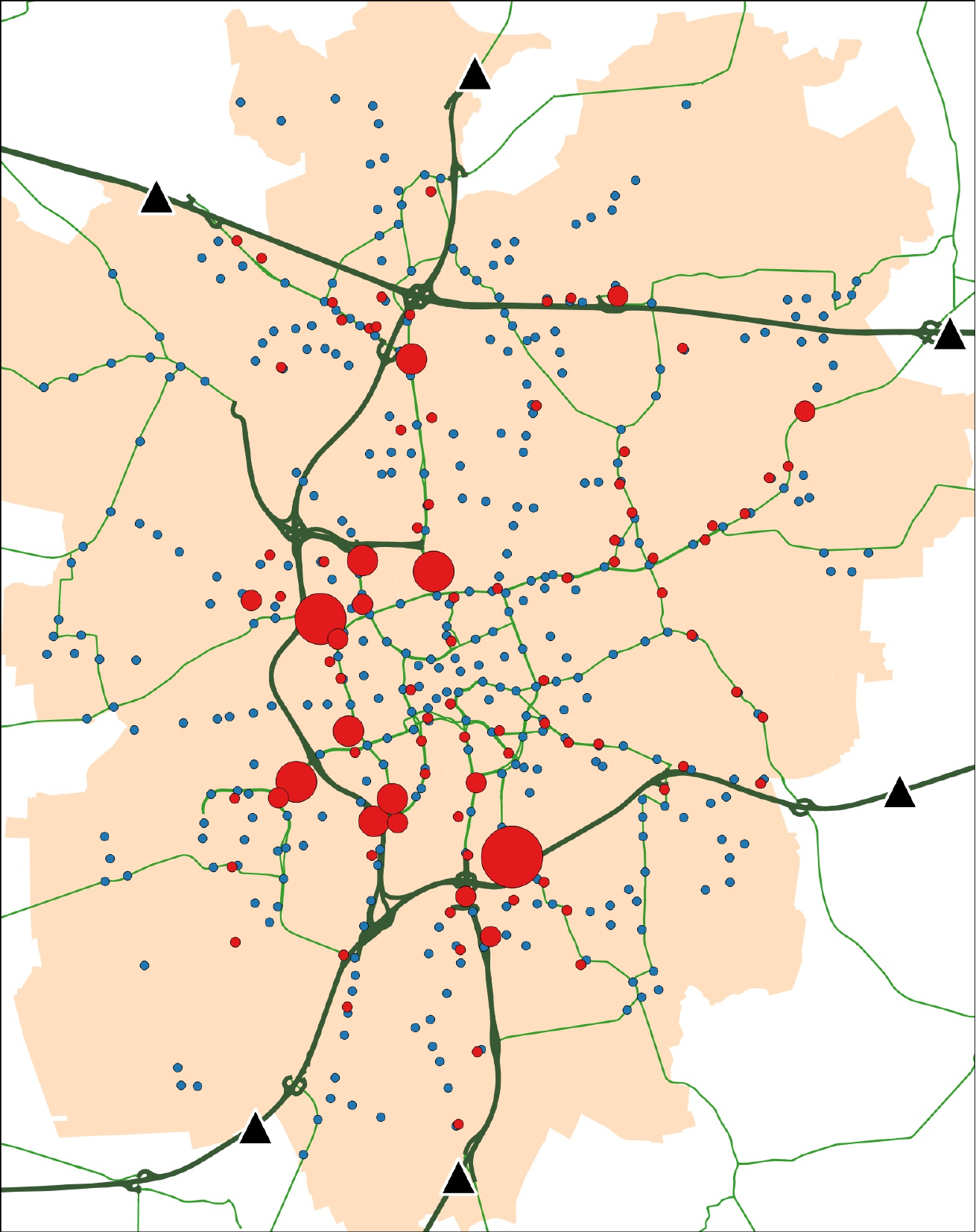} 
    \caption{MP selection frequency (Total)} 
    \label{subfig:sft3} 
    \vspace{1ex}
  \end{subfigure}
  \begin{subfigure}[b]{0.5\linewidth}
    \centering
    \includegraphics[width=0.4\linewidth]{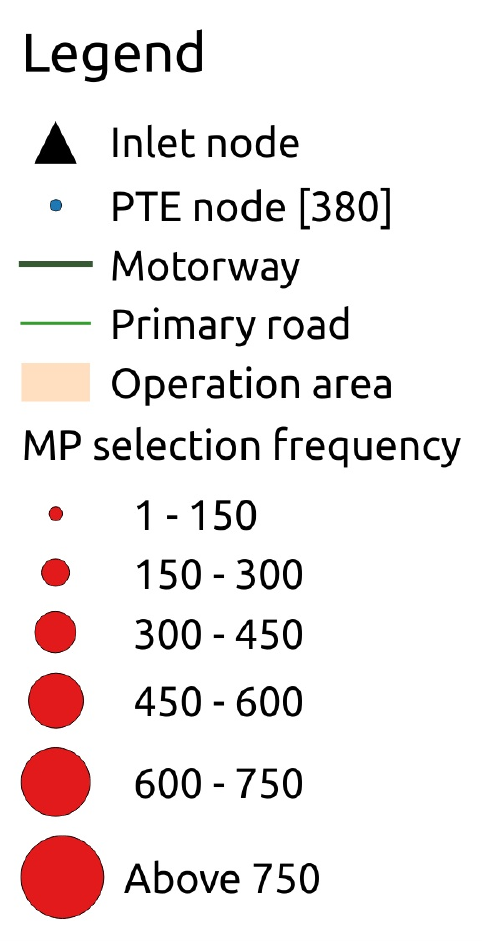} 
    \caption{Legend} 
    \label{subfig:sft4} 
    \vspace{1ex}
  \end{subfigure}

  \caption{Time-dependent meeting point selection frequency (10 000 simulation runs)}
  \label{fig:selection_frequency_times} 
\end{figure}

The first experiment (Figure \ref{fig:selection_frequency_times}) shows patterns for meeting point recommendations during noon at 12pm (Figure \ref{subfig:sft1}), in the later evening at 21am (Figure \ref{subfig:sft2}), and in total over the whole day (Figure \ref{subfig:sft3}). The maximum driver detour time was set to 30 minutes. The differences are only minor. Most changes between noon and evening hours can be explained by a reduced service level of public transportation.

\begin{figure}
  \begin{subfigure}[b]{0.5\linewidth}
    \centering
    \includegraphics[width=0.9\linewidth]{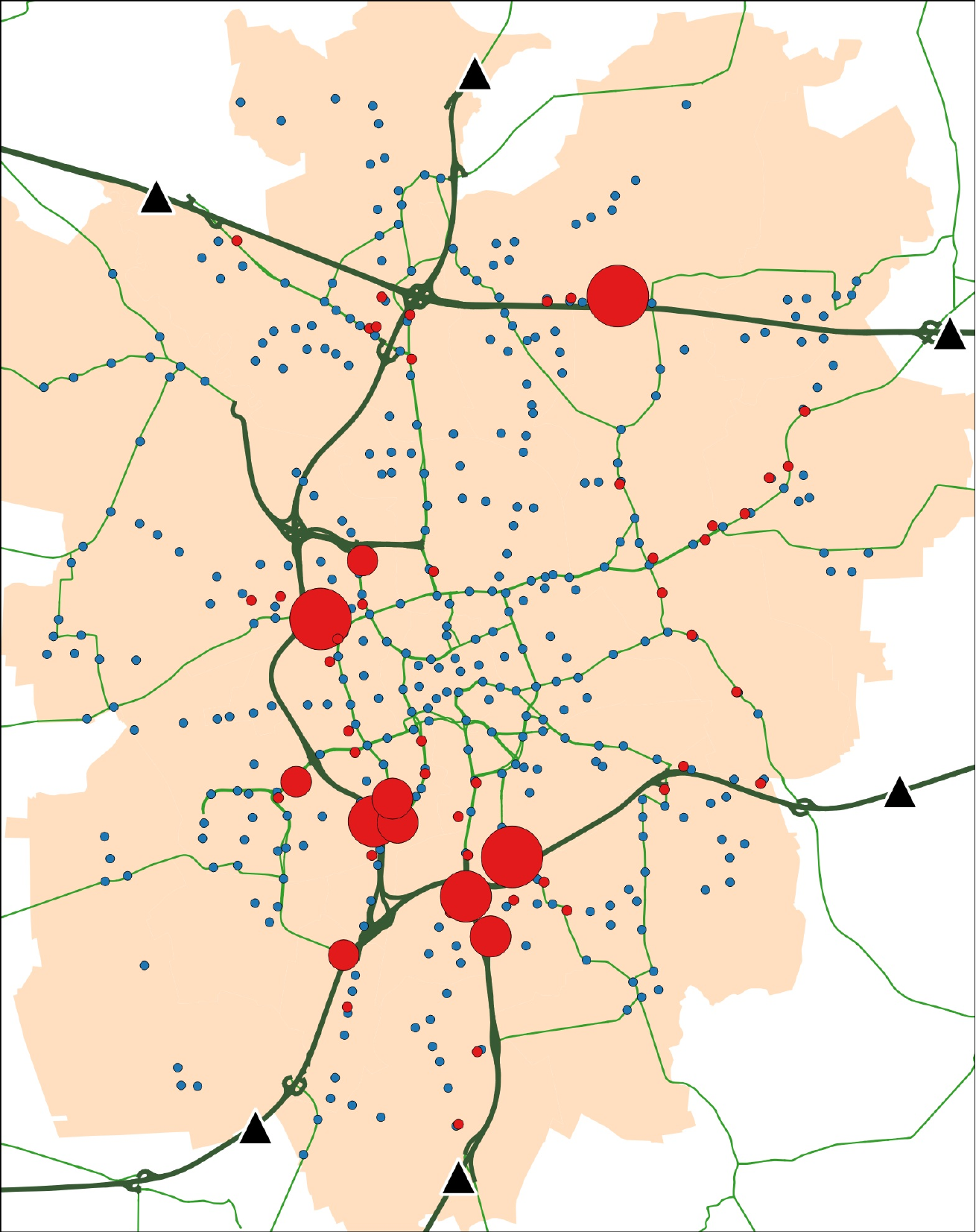} 
    \caption{MP selection frequency (5 minutes maximum driver detour)} 
    \label{subfig:sfd1} 
    \vspace{1ex}
  \end{subfigure}
  \begin{subfigure}[b]{0.5\linewidth}
    \centering
    \includegraphics[width=0.9\linewidth]{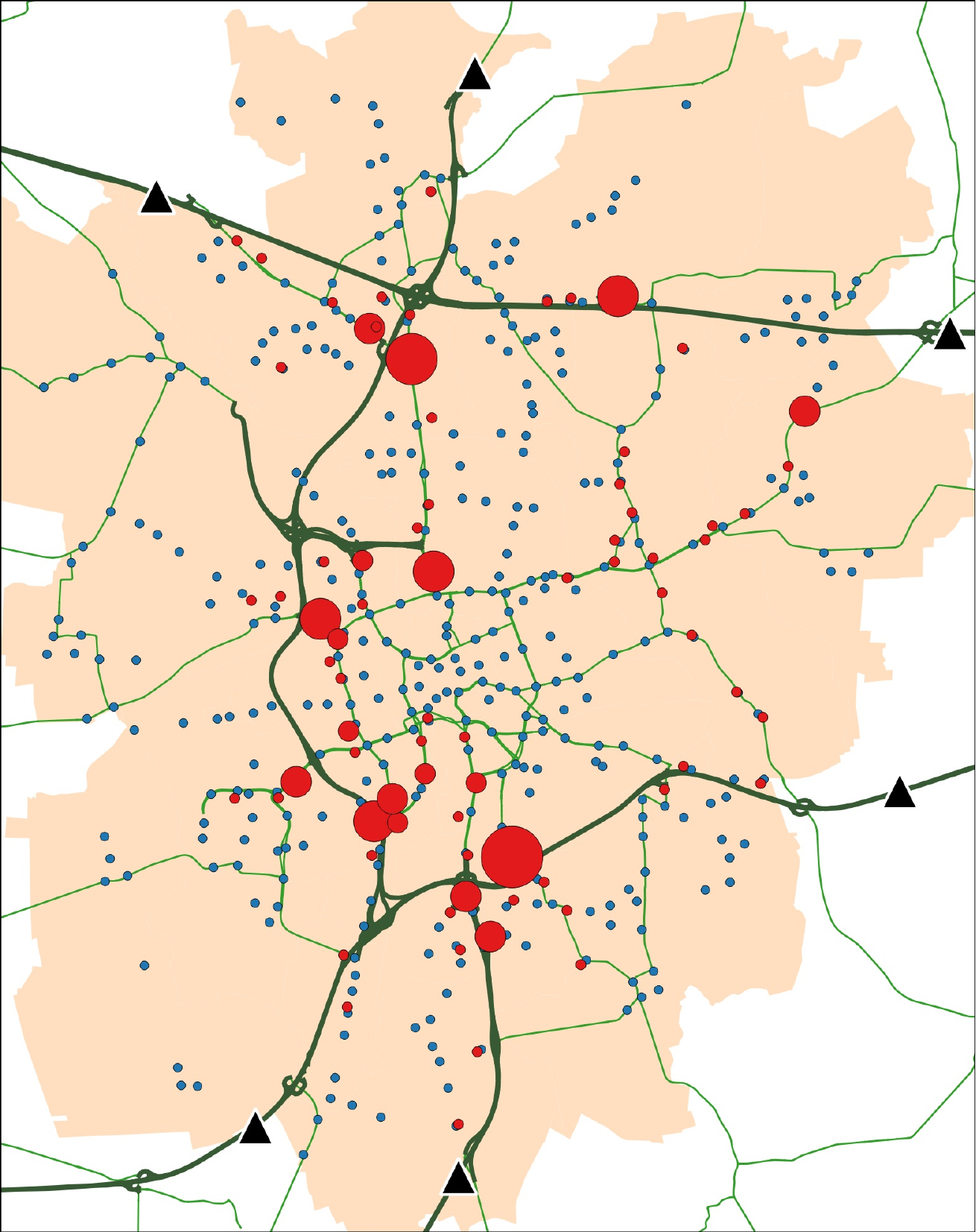} 
    \caption{MP selection frequency (10 minutes maximum driver detour)} 
    \label{subfig:sfd2} 
    \vspace{1ex}
  \end{subfigure}

  \caption{Detour-dependent meeting point selection frequency (10 000 simulation runs)}
  \label{fig:selection_frequency_detour} 
\end{figure}

\FloatBarrier
In a second experiment we varied the maximum allowed detour time parameter $t^{detr}_{*}$ that controls how far the drivers are willing or allowed to deviate from the direct route to reach a meeting point. Figure \ref{subfig:sfd1} shows the recommended meeting points for a tight 5 minutes threshold, Figure \ref{subfig:sfd2} for 10 minutes. Not surprisingly, the most frequently recommended meeting points are located very close to the motorway exit. In the 5 minutes case, the usage is very condensed into a few points. However, still some meeting points are selected far away from a motorway in the eastern part of the city. These points have been selected if a driver was taking the route from north east to south east (or reverse), and since there is no motorway, the route through the city is the shortest path anyway. 

\begin{figure}
  \begin{subfigure}[b]{0.5\linewidth}
    \centering
    \includegraphics[width=0.9\linewidth]{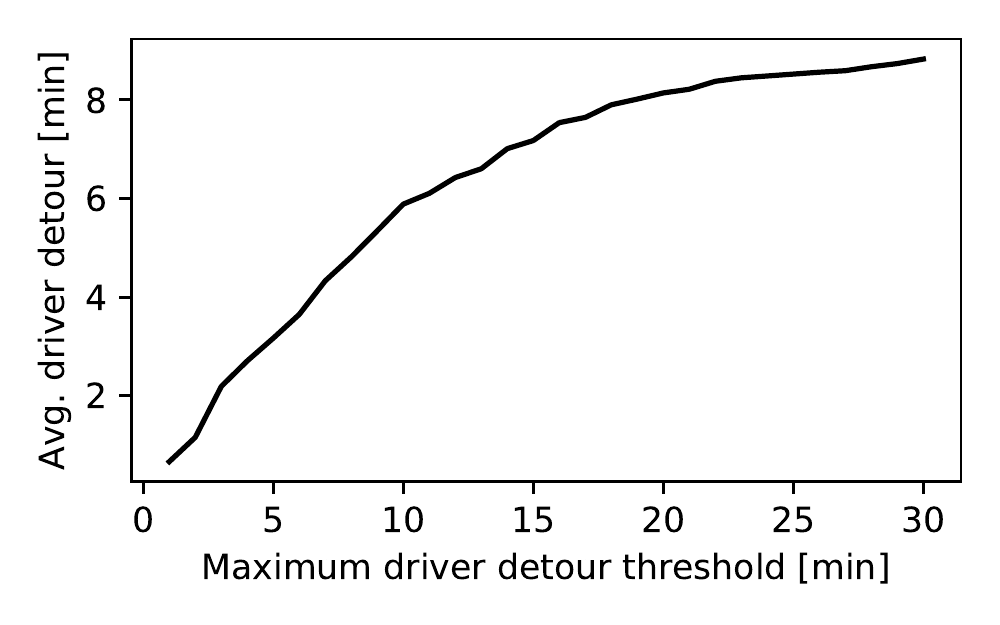} 
    \caption{Average driver detour time} 
    \label{subfig:smd1} 
    \vspace{1ex}
  \end{subfigure}
  \begin{subfigure}[b]{0.5\linewidth}
    \centering
    \includegraphics[width=0.9\linewidth]{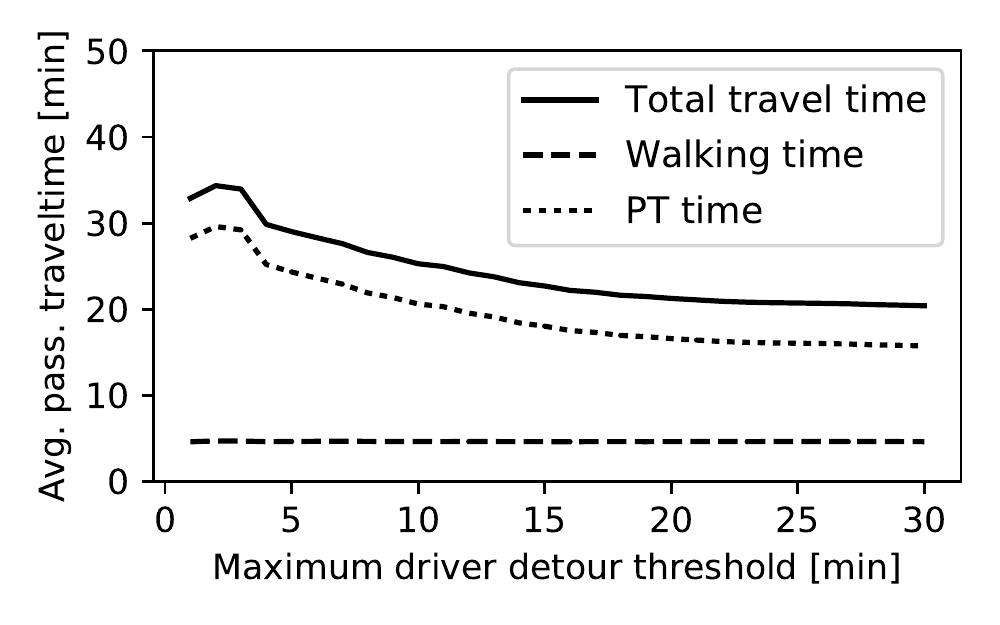} 
    \caption{Average passenger travel time} 
    \label{subfig:smd2} 
    \vspace{1ex}
  \end{subfigure}
  \newline
  \begin{subfigure}[b]{0.5\linewidth}
    \centering
    \includegraphics[width=0.9\linewidth]{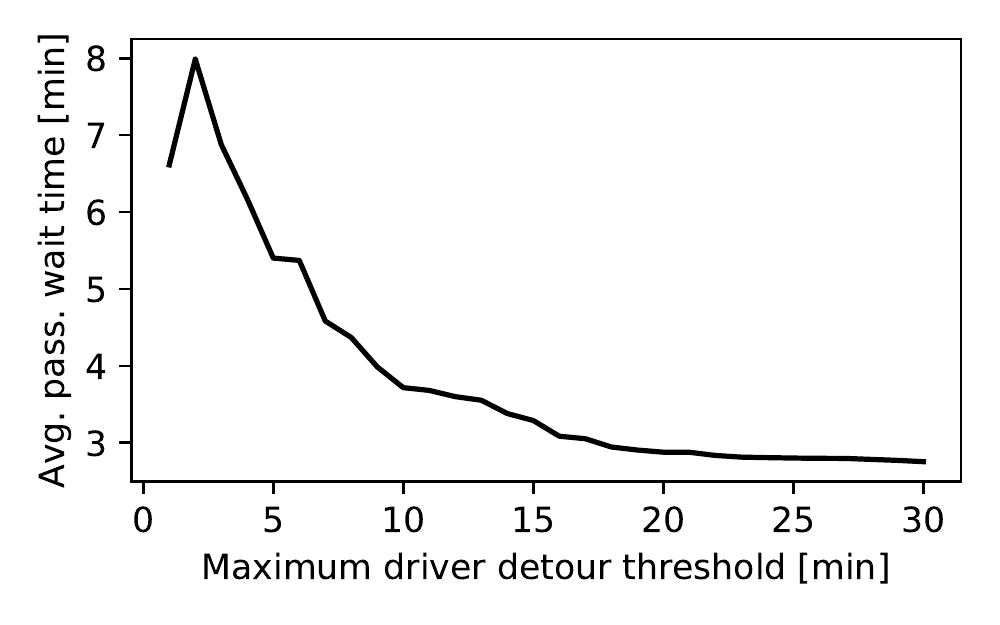} 
    \caption{Average passenger waiting time} 
    \label{subfig:smd3} 
    \vspace{1ex}
  \end{subfigure}
  \begin{subfigure}[b]{0.5\linewidth}
    \centering
    \includegraphics[width=0.9\linewidth]{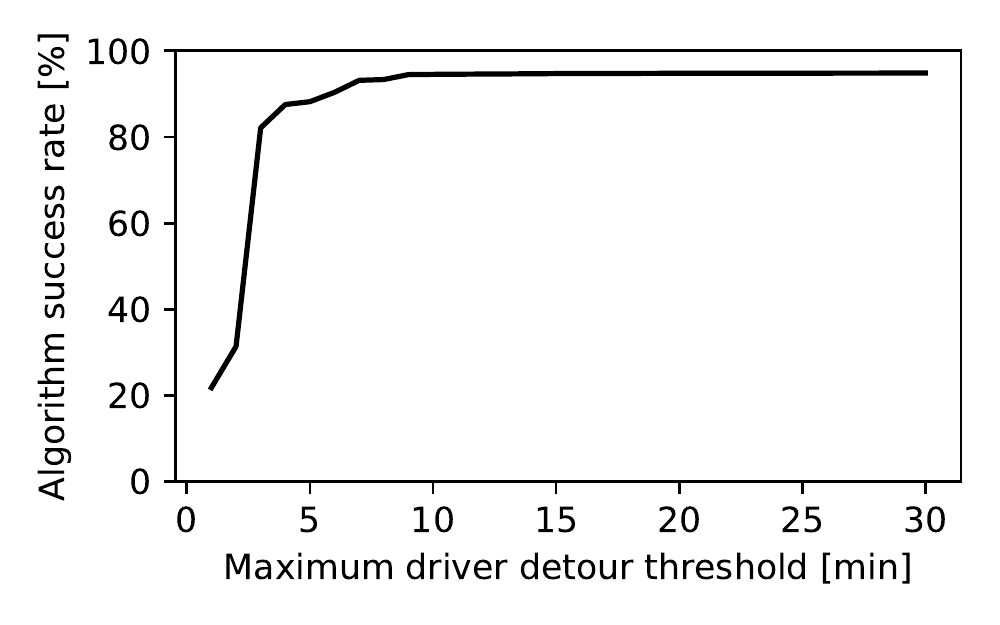} 
    \caption{Algorithm success rate} 
    \label{subfig:smd4} 
    \vspace{1ex}
  \end{subfigure}

  \caption{Simulation run with varying maximum detour threshold (10 000 runs each)}
  \label{fig:simulation_maxDetour} 
\end{figure}

Figure \ref{fig:simulation_maxDetour} shows various statistics when using different values for $t^{detr}_{*}$. Naturally, the average driver detour increases when the maximum allowed detour is increased, but not linearly (Figure \ref{subfig:smd1}). With no detour time restrictions, the average driver detour to reach a meeting point converges to a value between 8 and 9 minutes. The passengers have to travel less when the drivers are allowed or willing to deviate more from the direct route (Figure \ref{subfig:smd2}), which is again not surprising. 
The passenger waiting time is defined as the time that a customer has to wait at a meeting point for the driver because of an early arrival. Note that the waiting time is not included in the average passenger travel time. In Figure \ref{subfig:smd3} it can be seen that also the waiting time decreases with more flexibility in the meeting point selection. Finally, Figure \ref{subfig:smd4} shows the algorithm success rate, indicating how often the algorithm was able to find a valid solution satisfying all thresholds. As expected, with unreasonably strict detour thresholds (e.g. one minute), the algorithm is only able to find a solution in 20 \% of the runs. Already with 3 minutes detour, 80 \% of the requests can successfully be handled. 

As can also be seen, for 4.5 \% of the requests it is not possible to find a common meeting point satisfying the constraints at all, regardless of the threshold value. This happens because of requests from remote locations, where the public transport system is not offering rides frequently enough to reach the destination in time. In this case, the driver would have to manually negotiate a meeting point.

\FloatBarrier
The third experiment investigates the differences between using range voting and minimax voting, as described in section \ref{subsubsec:Voting}. The range voting leads to optimal overall time, i.e. a meeting point that reduces the total travel time of the group. In contrast, the minimax voting aims at selecting a meeting point that minimizes the maximum travel time of all participants. Table \ref{tab:votingCompare} shows the differences, based on different group sizes.

\begin{table*}[htb]
    \caption{Range voting vs. minimax voting (10 000 runs each)}
	\label{tab:votingCompare}
    \setlength{\tabcolsep}{6pt}
	\renewcommand{\arraystretch}{1.15}
	\centering
	\begin{tabular}{l|c c c c}
    \toprule[0.8pt]
       & 1 passenger & 2 passengers & 3 passengers & 4 passengers \\
    \toprule[0.8pt]
      Equal results of both voting rules & 59 \% & 31 \% & 30 \% & 29 \% \\
    \midrule[0.3pt]
      Different results: Average lateness \\using minimax voting & 12:29 min & 5:32 min & 4:30 min & 3:57 min \\
    \midrule[0.3pt]
      Different results: Average maximum \\travel time using range voting & 19:33 min & 34:04 min & 38:25 min & 41:14 min \\
    \midrule[0.3pt]
      Different results: Average maximum \\travel time using minimax voting & 9:20 min & 26:16 min & 31:11 min & 34:37 min \\
    \bottomrule[0.8pt]
    \end{tabular}
\end{table*}

The results show that the selection of a voting rule has a significant impact on the travel time and meeting point choice. If the driver meets with only one passenger, the same meeting point is recommended in 59 \% of the cases. With larger groups, this value decreases. If the voting results differ, the average maximum travel time can be approximately halved (from 19 to 9 minutes) when using minimax voting, at the cost of an enlarged total travel time of nearly 12 minutes. With more passengers involved, the maximum travel times increase in general, and the differences between the voting rules decrease.

\FloatBarrier
\section{Discussion}
\label{sec:Discussion}
The simulation shows that the proposed algorithm is, in theory, capable of handling many requests within a short time, and also the suggested locations are reasonable. Due to the precomputation, we could reach average processing response times of approximately 8 ms per request (Python 3.5 on Ubuntu Linux, running on AMD FX-6100 with 4 GB RAM), which makes real-time applications with a high request frequency conceivable. Of course, this comes at the cost of long precomputation times. For our network of Braunschweig it took approximately 10 hours (Python 3.5 and OpenTripPlanner instance on FreeBSD, running on Intel Xeon E5410 with 32 GB RAM). For a fully equipped real-world operation, the precomputation phase would have to be computed six times: standard weekday, Saturday and Sunday, for meeting points and for drop-off points. The storage space requirements are manageable with 17 MB in total for all saved connections using NumPy\footnote{NumPy is Python library (\url{http://www.numpy.org/})} binary format for storing matrices.

A limitation is clearly that the algorithm can only handle requests from drivers whose planned route passes an inlet node inbound and an inlet node outbound. Hence, drivers approaching from or leaving to more rural areas on smaller streets cannot be considered. However, it is often not a problem to insert more inlet nodes, since the driver time precomputation does not play a significant role compared to the public transport. Further, introducing new inlet points scales linearly.

Note further that the algorithm can only recommend one single meeting point, regardless of the passenger amount. In some cases it seems to be more reasonable to recommend two or more separate meeting points that have to be approached by the driver successively, e.g. if the first passenger is located close to a motorway exit on the west side, and the second passenger is located close to a motorway exit on the eastern side of the city. A simple workaround to tackle this issue is to apply the algorithm iteratively on the single passengers, if the necessary travel times for a common meeting point are assessed as unreasonably (or unacceptably) large, resulting in a recommendation of multiple different meeting points. The drawback is that the recommended meeting points are then not time optimal, since the driver travel times between the meeting points is not considered.

In our simulation, only a medium-sized city was investigated. If the service should be implemented for large metropolitan areas, it can be worth to split the area into smaller segments, in order to keep the amount of considered meeting points small. A hierarchical approach offers the possibility to search first for a feasible region, and then the actual meeting point can be determined for this area in detail in a second step.

Another advantage of the approach is that the algorithm is capable of reacting in real-time to congestion and disturbances in the street network. If such a change in driving times occurs, it is sufficient to add extra time to the corresponding values in the matrices $A^{inbound}_{\Psi}$ and $A^{outbound}_{\Psi}$, and the algorithm will automatically adapt to the modified situation. On the other hand, a drawback is that changes in the public transport network are more difficult to include. That means, when the route network or the timetable changes, the passenger travel time matrix $A^{P}$ has to be partly recalculated.

\FloatBarrier
\section{Conclusion}
\label{sec:Conclusion}
In this paper we introduced an algorithm to automatically recommend a meeting point to a driver/passenger group willing to meet in a city. While the driver uses a vehicle, the passengers are supposed to walk and/or use public transportation to reach the meeting point. The optimal point depends on the current location of the passengers as well as the time of driver arrival. The focus is on providing a real-time feasible solution with low response times. Hence, a precomputation of values is proposed. A ride sharing application can make use of the approach to suggest appropriate meeting points to long-distance ride-sharing customers. A simulation based on the proposed algorithm shows that, even with a high acceptable driver detour time, meeting points in the vicinity of motorway exits are frequently chosen, hence avoiding vehicle rides through inner city streets. Further, a maximum driver detour threshold can individually be applied, corresponding to the time budget of the driver.

\setcounter{secnumdepth}{0}
\section{Acknowledgements}
This research has been supported by the German Research Foundation (DFG) through the Research Training Group SocialCars (GRK 1931). The focus of the SocialCars Research Training Group is on significantly improving the city‘s future road traffic, through cooperative approaches. This support is gratefully acknowledged.

\bibliographystyle{tLBS}
\bibliography{ms}

\begin{thebibliography}{31}
\newcommand{\enquote}[1]{``#1''}
\providecommand{\natexlab}[1]{#1}
\providecommand{\url}[1]{\normalfont{#1}}
\providecommand{\urlprefix}{ }
\expandafter\ifx\csname urlstyle\endcsname\relax
  \providecommand{\doi}[1]{doi:\discretionary{}{}{}#1}\else
  \providecommand{\doi}{doi:\discretionary{}{}{}\begingroup
  \urlstyle{rm}\Url}\fi
\providecommand{\eprint}[2][]{\url{#2}}

\bibitem[Agatz et~al.(2012)Agatz, Erera, Savelsbergh, and
  Wang]{agatz2012optimization}
Agatz, Niels, Alan Erera, Martin Savelsbergh, and Xing Wang. 2012.
  ``Optimization for dynamic ride-sharing: A review.'' \emph{European Journal
  of Operational Research} 223 (2): 295--303.

\bibitem[Agatz et~al.(2011)Agatz, Erera, Savelsbergh, and
  Wang]{agatz2011dynamic}
Agatz, Niels~AH, Alan~L Erera, Martin~WP Savelsbergh, and Xing Wang. 2011.
  ``Dynamic ride-sharing: A simulation study in metro Atlanta.''
  \emph{Transportation Research Part B: Methodological} 45 (9): 1450--1464.

\bibitem[Aissat and Oulamara(2014)]{aissat2014dynamic}
Aissat, Kamel, and Ammar Oulamara. 2014. ``Dynamic ridesharing with
  intermediate locations.'' In \emph{Computational Intelligence in Vehicles and
  Transportation Systems (CIVTS), 2014 IEEE Symposium on,} 36--42. IEEE.

\bibitem[Aissat and Oulamara(2015)]{aissat2015meeting}
Aissat, K, and A~Oulamara. 2015. ``Meeting Locations in Real-Time Ridesharing
  Problem: A Buckets Approach.'' In \emph{Operations Research and Enterprise
  Systems,} 71--92. Springer.

\bibitem[A{\"\i}vodji et~al.(2016)A{\"\i}vodji, Gambs, Huguet, and
  Killijian]{aivodji2016meeting}
A{\"\i}vodji, Ulrich~Matchi, S{\'e}bastien Gambs, Marie-Jos{\'e} Huguet, and
  Marc-Olivier Killijian. 2016. ``Meeting points in ridesharing: A
  privacy-preserving approach.'' \emph{Transportation Research Part C: Emerging
  Technologies} 72: 239--253.

\bibitem[Balardino and Santos(2016)]{balardino2016heuristic}
Balardino, Allan~F, and Andr{\'e}~G Santos. 2016. ``Heuristic and Exact
  Approach for the Close Enough Ridematching Problem.'' In \emph{Hybrid
  Intelligent Systems,} 281--293. Springer.

\bibitem[Bast et~al.(2015)Bast, Delling, Goldberg, M{\"u}ller-Hannemann, Pajor,
  Sanders, Wagner, and Werneck]{bast2015route}
Bast, Hannah, Daniel Delling, Andrew Goldberg, Matthias M{\"u}ller-Hannemann,
  Thomas Pajor, Peter Sanders, Dorothea Wagner, and Renato~F Werneck. 2015.
  ``Route planning in transportation networks.'' \emph{arXiv preprint
  arXiv:1504.05140} .

\bibitem[{BlaBlaCar}(2017)]{BlaBlaCarMP}
{BlaBlaCar}. 2017. ``{Ridesharing Meeting Points}.''
  \url{https://www.blablacar.com/blog/blablalife/travel-tips/ridesharing-meeting-points}.
  [Online; accessed 01.06.2017].

\bibitem[Cai, Kloks, and Wong(1997)]{cai1997time}
Cai, X, Ton Kloks, and Chak-Kuen Wong. 1997. ``Time-varying shortest path
  problems with constraints.'' \emph{Networks} 29 (3): 141--150.

\bibitem[Chabini(1998)]{chabini1998discrete}
Chabini, Ismail. 1998. ``Discrete dynamic shortest path problems in
  transportation applications: Complexity and algorithms with optimal run
  time.'' \emph{Transportation Research Record: Journal of the Transportation
  Research Board}  (1645): 170--175.

\bibitem[Czioska, Mattfeld, and Sester(2017)]{Czioska2017314}
Czioska, Paul, Dirk~C. Mattfeld, and Monika Sester. 2017. ``GIS-based
  identification and assessment of suitable meeting point locations for
  ride-sharing.'' \emph{Transportation Research Procedia} 22: 314 -- 324.
  \urlprefix\url{http://www.sciencedirect.com/science/article/pii/S2352146517301734}.

\bibitem[Demiryurek, Banaei-Kashani, and Shahabi(2010)]{demiryurek2010case}
Demiryurek, Ugur, Farnoush Banaei-Kashani, and Cyrus Shahabi. 2010. ``A case
  for time-dependent shortest path computation in spatial networks.'' In
  \emph{Proceedings of the 18th SIGSPATIAL International Conference on Advances
  in Geographic Information Systems,} 474--477. ACM.

\bibitem[Dennisen and M{\"u}ller(2015)]{dennisen2015agent}
Dennisen, Sophie~L, and J{\"o}rg~P M{\"u}ller. 2015. ``Agent-Based Voting
  Architecture for Traffic Applications.'' In \emph{German Conference on
  Multiagent System Technologies,} 200--217. Springer.

\bibitem[Ester et~al.(1996)Ester, Kriegel, Sander, Xu et~al.]{ester1996density}
Ester, Martin, Hans-Peter Kriegel, J{\"o}rg Sander, Xiaowei Xu, et~al. 1996.
  ``A density-based algorithm for discovering clusters in large spatial
  databases with noise..'' In \emph{Kdd,}  Vol.~96226--231.

\bibitem[Furuhata et~al.(2013)Furuhata, Dessouky, Ord{\'o}{\~n}ez, Brunet,
  Wang, and Koenig]{furuhata2013ridesharing}
Furuhata, Masabumi, Maged Dessouky, Fernando Ord{\'o}{\~n}ez, Marc-Etienne
  Brunet, Xiaoqing Wang, and Sven Koenig. 2013. ``Ridesharing: The
  state-of-the-art and future directions.'' \emph{Transportation Research Part
  B: Methodological} 57: 28--46.

\bibitem[Goel, Kulik, and Ramamohanarao(2016)]{goel2016privacy}
Goel, Preeti, Lars Kulik, and Kotagiri Ramamohanarao. 2016. ``Privacy-Aware
  Dynamic Ride Sharing.'' \emph{ACM Transactions on Spatial Algorithms and
  Systems} 2 (1): 4.

\bibitem[H{\"a}gerstraand(1970)]{hagerstraand1970people}
H{\"a}gerstraand, Torsten. 1970. ``What about people in regional science?.''
  \emph{Papers in regional science} 24 (1): 7--24.

\bibitem[INFAS and DLR(2008)]{infas2010mobilitat}
INFAS, and DLR. 2008. ``{Mobilit{\"a}t in Deutschland 2008}.''
  \url{http://www.mobilitaet-in-deutschland.de}. [Online; accessed 01.04.2016].

\bibitem[Maue, Sanders, and Matijevic(2009)]{maue2009goal}
Maue, Jens, Peter Sanders, and Domagoj Matijevic. 2009. ``Goal-directed
  shortest-path queries using precomputed cluster distances.'' \emph{Journal of
  Experimental Algorithmics (JEA)} 14: 2.

\bibitem[Miller(1991)]{miller1991modelling}
Miller, Harvey~J. 1991. ``Modelling accessibility using space-time prism
  concepts within geographical information systems.'' \emph{International
  Journal of Geographical Information System} 5 (3): 287--301.

\bibitem[M{\"u}ller-Hannemann et~al.(2007)M{\"u}ller-Hannemann, Schulz, Wagner,
  and Zaroliagis]{muller2007timetable}
M{\"u}ller-Hannemann, Matthias, Frank Schulz, Dorothea Wagner, and Christos
  Zaroliagis. 2007. ``Timetable information: Models and algorithms.'' In
  \emph{Algorithmic Methods for Railway Optimization,} 67--90. Springer.

\bibitem[Orda and Rom(1990)]{orda1990shortest}
Orda, Ariel, and Raphael Rom. 1990. ``Shortest-path and minimum-delay
  algorithms in networks with time-dependent edge-length.'' \emph{Journal of
  the ACM (JACM)} 37 (3): 607--625.

\bibitem[Pyrga et~al.(2008)Pyrga, Schulz, Wagner, and
  Zaroliagis]{pyrga2008efficient}
Pyrga, Evangelia, Frank Schulz, Dorothea Wagner, and Christos Zaroliagis. 2008.
  ``Efficient models for timetable information in public transportation
  systems.'' \emph{Journal of Experimental Algorithmics (JEA)} 12: 2--4.

\bibitem[Raubal et~al.(2007)Raubal, Winter, Te$\beta$mann, and
  Gaisbauer]{raubal2007time}
Raubal, Martin, Stephan Winter, Sven Te$\beta$mann, and Christian Gaisbauer.
  2007. ``Time geography for ad-hoc shared-ride trip planning in mobile
  geosensor networks.'' \emph{ISPRS Journal of Photogrammetry and Remote
  Sensing} 62 (5): 366--381.

\bibitem[Rigby, Kr\"{u}ger, and Winter(2013)]{Rigby:2013:OCU:2525314.2525334}
Rigby, Michael, Antonio Kr\"{u}ger, and Stephan Winter. 2013. ``An
  Opportunistic Client User Interface to Support Centralized Ride Share
  Planning.'' In \emph{Proceedings of the 21st ACM SIGSPATIAL International
  Conference on Advances in Geographic Information Systems,} Orlando, Florida.
  34--43. ACM.

\bibitem[Rigby and Winter(2015)]{rigby2015enhancing}
Rigby, Michael, and Stephan Winter. 2015. ``Enhancing launch pads for
  decision-making in intelligent mobility on-demand.'' \emph{Journal of
  Location Based Services} 9 (2): 77--92.

\bibitem[Rothe et~al.(2012)Rothe, Baumeister, Lindner, and
  Rothe]{rothe2012einfuhrung}
Rothe, J{\"o}rg, Dorothea Baumeister, Claudia Lindner, and Irene Rothe. 2012.
  \emph{Einf{\"u}hrung in Computational Social Choice: Individuelle Strategien
  und kollektive Entscheidungen beim Spielen, W{\"a}hlen und Teilen}.
  Springer-Verlag.

\bibitem[Sanders and Schultes(2005)]{sanders2005highway}
Sanders, Peter, and Dominik Schultes. 2005. ``Highway hierarchies hasten exact
  shortest path queries.'' In \emph{European Symposium on Algorithms,}
  568--579. Springer.

\bibitem[Stiglic et~al.(2015)Stiglic, Agatz, Savelsbergh, and
  Gradisar]{stiglic2015benefits}
Stiglic, Mitja, Niels Agatz, Martin Savelsbergh, and Mirko Gradisar. 2015.
  ``The benefits of meeting points in ride-sharing systems.''
  \emph{Transportation Research Part B: Methodological} 82: 36--53.

\bibitem[{U.S. Department of Transportation}(2015)]{dot2015national}
{U.S. Department of Transportation}. 2015. ``{National transportation
  statistics}.'' \url{http://www.rita.dot.gov/bts/data_and_statistics}.
  [Online; accessed 01.04.2016].

\bibitem[Winter and Nittel(2006)]{winter2006ad}
Winter, Stephan, and Silvia Nittel. 2006. ``Ad hoc shared-ride trip planning by
  mobile geosensor networks.'' \emph{International Journal of Geographical
  Information Science} 20 (8): 899--916.

\end{thebibliography}

\newpage
\section{Algorithms}
\label{algs}

\begin{algorithm}
    \begin{algorithmic}
	\caption{Algorithm used to sample stop positions across service area}
	\State Given: List of PTE nodes $\Pi$, Threshold $d_{*}$
    \For{$i \in \{0,1,\cdots,|\Pi|\}$}
	    \State $\Lambda \gets $ \Call{k-Means}{data=$\Pi$, clusterCount=$i$} \Comment{Get cluster center positions}
        \For{$\pi \in \Pi$} \Comment{Iterate through all PTE nodes}
	        \If{$\nexists \: \lambda \in \Lambda \: | \: $\Call{Dist}{$\pi, \lambda$} $\leq d_{*}$}
			    \State Break inner loop and continue outer loop
		    \EndIf
        \EndFor
        \State \Return $\Lambda$ \Comment{All PTE nodes are covered within certain distance}
	\EndFor
	\end{algorithmic}
\end{algorithm}

\begin{algorithm}
   \begin{algorithmic}
	\caption{Meeting point selection with clustering}
	\State Given: Meeting point candiate (MPC) set $M$, Threshold $d^{gap}_{*}$
	\State Initialize result set $\Phi \gets \{ \}$
	\State Initialize cluster check set $\Theta \gets \{ \}$
	\State $C(M) \gets $\Call{DBSCAN}{data=$M$, threshold=$d^{gap}_{*}$} \Comment{MPC assignments to clusters}
	\State $U(M) \gets $\Call{DetermineUsage}{$M$} \Comment{Frequency of MPCs being selected}
	\For{$m \in $ \Call{Sort}{$M,U$}} \Comment{Sort MPCs by descending selection frequency}
	    \If{$C(m) \in \Theta$}
			\If{(\Call{Dist}{$m,n$} $\geq d^{gap}_{*}) \: \forall \: n \in \{ \Phi \: | \: C(n) = C(m) \}$}
				\State $\Phi \gets \Phi \cup m$
			\EndIf
		\Else
			\State $\Phi \gets \Phi \cup m$
			\State $\Theta \gets \Theta \cup C(m)$
		\EndIf
	\EndFor
	\State \Return $\Phi$
	\end{algorithmic}
\end{algorithm}

\label{lastpage}
\end{document}